\documentclass[useAMS,usenatbib]{mn2e}

 \usepackage{times}

\usepackage{graphicx}
\usepackage{rotating}
\usepackage{lscape}
\usepackage{epsfig}
\usepackage{amssymb}
\usepackage{myaasmacros}
\usepackage{amsmath}
\usepackage{multicol}
\usepackage{color}

\def\lesssim{\mathrel{\hbox{\rlap{\hbox{\lower4pt\hbox{$\sim$}}}\hbox{$<$}}}}
\def\gtrsim{\mathrel{\hbox{\rlap{\hbox{\lower4pt\hbox{$\sim$}}}\hbox{$>$}}}}
\def\subsun{\mbox{$_{\normalsize\odot}$}}
\def\arcsec{\hbox{$^{\prime\prime}$}\,}
\def\arcmin{$^{\prime}$\,}
\def\deg{\hbox{$^\circ$}}
\def\mic{$\,\mu $m\,}

\def\gsim{\mathrel{\lower0.6ex\hbox{$\buildrel {\textstyle >}
\over {\scriptstyle \sim}$}}}
\def\lsim{\mathrel{\lower0.6ex\hbox{$\buildrel {\textstyle <}
\over {\scriptstyle \sim}$}}}

\title[Herschel overdensities around HzRG]{Searching for large--scale structures around high-redshift radio galaxies with {\it Herschel}}
\author[E.E. Rigby]{\parbox{\textwidth}{E.~E. Rigby$^{1}$\thanks{E-mail: emmaerigby@gmail.com}, N.~A. Hatch$^{2}$, H.~J.~A. R\"{o}ttgering$^{1}$, B. Sibthorpe$^{3}$, Y.~K. Chiang$^{4}$, R. Overzier$^{4,5}$,  
R. Herbonnet$^{1}$, 
S. Borgani$^{6}$, 
D.~L. Clements$^{7}$, 
H. Dannerbauer$^{8}$, 
C. De Breuck$^{9}$, 
G. De Lucia$^{6}$, 
J. Kurk$^{10}$, 
F. Maschietto$^{1}$, 
G. Miley$^{1}$, 
A. Saro$^{11}$, 
N. Seymour$^{12}$, 
B. Venemans$^{13}$
\\}\\
$^{1}$Leiden Observatory, P.O. Box 9513, 2300 RA, Leiden, The Netherlands\\
$^{2}$School of Physics and Astronomy, University of Nottingham, University Park, Nottingham NG7 2RD, UK\\
$^{3}$SRON Netherlands Institute for Space Research, Landleven 12, 9747 AD, Groningen, The Netherlands\\
$^{4}$Department of Astronomy, The University of Texas at Austin, Austin, TX 78712\\
$^{5}$Observat\'orio Nacional, Rua Jos\'e Cristino, 77. CEP 20921-400, S\~ao Crist\'ov\~ao, Rio de Janeiro-RJ, Brazil\\
$^{6}$INAF - Osservatorio Astronomico di Trieste, Via G.B. Tiepolo 11, I-34143 Trieste, Italy\\
$^{7}$Astrophysics Group, Imperial College London, Blackett Laboratory, Prince Consort Road, London SW7 2AZ, UK\\
$^{8}$Institut f\"{u}r Astrophysik, Universit\"{a}t Wien, T\"{u}rkenschanzstra\ss{}e 17, A-1180 Wien, Austria\\
$^{9}$European Southern Observatory, Karl-Schwarzschild-Stra\ss{}e 2, D-85748 Garching bei M\"{u}nchen, Germany\\
$^{10}$Max-Planck-Institut f\"{u}r Extraterrestrische Physik, Postfach 1312, Giessenbachstrasse, D-85741 Garching, Germany\\
$^{11}$Department of Physics, Ludwig-Maximilians-Universit\"{a}t, Scheinerstr. 1, D-81679 M\"{u}nchen, Germany\\
$^{12}$CSIRO Astronomy and Space Science, PO Box 76, Epping, NSW 1710, Australia\\
$^{13}$Max-Planck Institute for Astronomy, K\"{o}nigstuhl 17, D-69117 Heidelberg, Germany
}
\begin{document}

\date{}

\pagerange{\pageref{firstpage}--\pageref{lastpage}} \pubyear{2011}

\maketitle

\label{firstpage}

\begin{abstract}
This paper presents the first results of a far--infrared search for protocluster--associated galaxy overdensities using the SPIRE instrument on--board the {\it Herschel} Space Observatory. Large ($\sim$400 arcmin$^{2}$) fields surrounding 26 powerful high--redshift radio galaxies ($2.0 < z < 4.1$; $L_{\rm 500 MHz} > 10^{28.5}$ WHz$^{-1}$) are mapped at 250, 350 and 500\mic to give a unique wide--field sample.

On average the fields have a higher than expected, compared to blank fields, surface density of 500\mic sources within 6 comoving Mpc of the radio galaxy. The analysis is then restricted to potential protocluster members only, which are identified using a far--infrared colour selection; this reveals significant overdensities of galaxies in 2 fields, neither of which are previously known protoclusters. The probability of finding 2 overdensities of this size by chance, given the number of fields observed is $5 \times 10^{-4}$. Overdensities here exist around radio galaxies with $L_{\rm 500 MHz} \gtrsim 10^{29}$ WHz$^{-1}$ and $z < 3$. The radial extent of the average far--infrared overdensity is found to be $\sim$6 comoving Mpc. 

Comparison with predictions from numerical simulations shows that the overdensities are consistent with having masses $> 10^{14}$M$_{\subsun}$. However, the large uncertainty in the redshift estimation means that it is possible that these far--infrared overdensities consist of several structures across the redshift range searched.

\end{abstract}


\section{Introduction}

Protoclusters, the high--redshift ancestors of local galaxy clusters, are powerful laboratories both for tracing the emergence of large scale structure and for studying the evolution of galaxies in dense environments. Observations of these structures over a range of wavelengths can map both the star forming and evolved galaxy populations, facilitating studies of the build-up of galaxies and clusters during a crucial epoch of their evolution. 

Luminous high-redshift radio galaxies (HzRGs) are among the most massive galaxies in the early Universe, and likely progenitors of Brightest Cluster Galaxies (BCGs); for a review see \citet{miley2008}. They are often embedded in overdense structures of galaxies with sizes $>$3 Mpc and present--day masses in the range 10$^{14}$--10$^{15}$M$_{\odot}$, comparable to local rich clusters \citep[e.g.][]{venemans2007, hatch2011}. Targeting them has therefore proven to be an efficient technique for identifying these overdense structures in the early Universe. The statistics of radio galaxy environment luminosity functions are consistent with every brightest cluster galaxy having gone through a `HzRG' evolutionary phase, with radio-selected protoclusters being typical ancestors of local galaxy clusters \citep[][2007]{venemans2002}. 

As protoclusters evolve into today's massive clusters, galaxies and gas are fed into the central regions from the surrounding environment. Therefore to study the large-scale structure whose collapse gives rise to local virialized clusters, scales extending well beyond the protocluster cores at $z > 2$ must be mapped. A typical galaxy, travelling at 500 kms$^{-1}$, will cross $\sim$5 Mpc in the time between $z = 3$ and the present day. So the material that ends up within 5 Mpc of the cluster core will be spread across $\sim$20 comoving Mpc at z $\sim$3. \citet{intema} studied Lyman-break galaxies in the 15 Mpc environment around a $z = 4.1$ protocluster, and found that the central protocluster concentration was surrounded by several interconnected galaxy concentrations, similar to the predictions of cosmological simulations. Similarly \citet{hayashi} studied H$\alpha$ emitters surrounding a $z \sim 2.5$ protocluster and found that they lie in 3 distinct clumps, spread over $\sim$8 comoving Mpc.

Far-infrared observations provide a unique opportunity to find and characterise distant star-forming cluster members. Cluster ellipticals are an important and enigmatic population; one of the major outstanding mysteries of galaxy formation is how these massive, `red and dead' galaxies grow quickly enough to be fully-formed by $z=1$. A key unknown ingredient is the amount of dust-obscured star formation that occurs at $2 < z < 4$, when these galaxies build-up most of their stellar mass. This is impossible to determine without far-infrared data to break the degeneracy between age and dust content when modelling the UV to infrared spectral energy distribution (SED). Previous detections of far--infrared/sub--millimetre galaxy overdensities with ground--based instruments such as SCUBA \citep{stevens} and MAMBO \citep{breuck} have shown that this wavelength regime is promising for protocluster searches.

Historically the normal field of view of most infra--red or optical imagers is comparable, or even smaller than the predicted angular size of protoclusters ($\sim$20\arcmin), which has severely hampered their study. This situation has improved with the advent of the {\it Herschel} Space Observatory \protect\citep{herschel}\footnote{Herschel is an ESA space observatory with science instruments provided by European-led Principal Investigator consortia and with important participation from NASA.}, with its combination of sensitivity and fast mapping speed. Far--infrared protocluster candidates are beginning to be discovered within the surveys carried out with {\it Herschel}. \citet{ivison} spectroscopically identified a cluster of star--bursting proto--elliptical galaxies at $z=2.41$ in the {\it Herschel}--ATLAS survey \citep{eales}. A similar detection of an overdensity of hyper--luminous infrared galaxies at $z = 3.3$ (Clements et al. in prep) was made by cross–matching the ATLAS catalogue with the Planck Early Release Compact Source catalogue \citep{herranz}, while a similar cross identification with HerMES sources reveals four lower redshift cluster candidates (Clements et al., 2013, MNRAS submitted). However, protoclusters are rare structures so these surveys are not likely to contain many.

The need for a large--area, targeted, survey of HzRG fields has motivated the development of the Search for Protoclusters with {\it Herschel} (SPHer), a far-infrared survey of 26 HzRG fields. This takes advantage of the combination of sensitivity and fast mapping speed of the Spectral and Photometric Imaging REciever \citep[SPIRE;][]{spire} on--board {\it Herschel}  to map the entire protocluster and its environment. This first paper uses the SPIRE data from the 26 fields to investigate the far-infrared environments around the HzRGs. Future papers will address the multiwavelength properties of these fields.

The layout of the paper is as follows: Sections \ref{spher} and \ref{dr} describe the SPHer survey, observations and data reduction; Section \ref{cat} outlines the creation of the corresponding source catalogues and Section \ref{env} analyses the central far--infrared environments of each field. Finally Sections \ref{discuss} and \ref{conclude} discus the results. Throughout this paper a concordance cosmology is assumed, with values for the cosmological parameters of H$_{0} = 71$~km~s$^{-1}$Mpc$^{-1}$, $\Omega_{\rm m} = 0.27$ and $\Omega_{\rm
  \Lambda} = 0.73$ \citep{spergel}. This gives a physical scale at $z=2$ of 8.475 kpc$/$\arcsec and 7.083 kpc$/$\arcsec at $z=4$. These correspond to $\sim$1.5 comoving Mpc$/$\arcmin and $\sim$2.1 comoving Mpc$/$\arcmin respectively.

\section{The {\it Herschel} protocluster survey}
\label{spher}

The aim of the SPHer survey was to use the far-infrared SPIRE instrument to construct a statistical sample of HzRG fields hosting the most massive black holes known at $2 < z < 4$, across an area large enough to encompass the entire potential protocluster and its environment. This means that sufficient numbers of these objects are needed at each epoch to ensure against bias by atypical targets, and provide statistically robust results. The compendium of \citet{miley2008} was used to select 26 of the most powerful HzRGs ($L_{\rm 500 MHz} > 10^{28.5}$ W Hz$^{-1}$), evenly spaced across this redshift range, with a minimum of 2 sources per redshift bin, $\Delta z = 0.5$. Eight of these targets are already confirmed as rich protoclusters (via Ly$\alpha$ or H$\alpha$ observations), and possess a large amount of supporting data spanning 0.3 to 24\mic, though this generally covers the inner ($< 5$\arcmin) region only.

The SPHer far-infrared observations were carried out as part of the first (OT1) and second (OT2) Open Time {\it Herschel} programmes. The initial OT1 proposal mapped 30\arcmin $\times$ 30\arcmin regions, centred on the central HzRG, in 7 of the confirmed protoclusters, and one protocluster candidate; the remaining 18 fields were targeted in the OT2 period over a smaller area of 20\arcmin $\times$ 20\arcmin to reduce the observing time requirements. Table \ref{obs_table} lists the full set of HzRG targets, together with their redshifts, radio powers and the {\it Herschel} observing programme within which they fall.

\begin{table*}
\begin{tabular}{llccccccc}
\hline
Label & Name & RA & Dec & z & $\log$ L$_{\rm 500 MHz}$& {\it Herschel}  & OBSIDs & Protocluster \\
     & & (J2000) & (J2000) & & (W Hz$^{-1}$)&Programme & & reference\\
\hline
1   & MPJ 1758-6738	     & 17:58:52.9 & -67:38:34   & 2.03  & 28.99 & OT2 &  1342241076 - 1342241080 &\\  
2   & MRC 1138-262         & 11:40:48.3 & -26:29:09   & 2.16   & 29.07 & OT1   &  1342248476 - 1342248480& \citet{kurk2000}\\
3   & BRL 0128-264         & 01:30:27.9 & -26:09:57   & 2.35  & 29.13 & OT2   &  1342258394 - 1342258398&\\
4   & USS 1707+105         & 17:10:07   & +10:31:06   & 2.35   & 28.63 & OT1   &  1342227649 - 1342227653& \citet{hatch2011}\\
5   & MRC 0406-244	     & 04:08:51.4 & -24:18:17   & 2.44   & 29.03 & OT2 &  1342241117 - 1342241121&\\
6   & MG 2308+0336         & 23:08:25.1 & +03:37:04   & 2.46   & 28.51 & OT1   &  1342211611	         & \\
7   & 3C 257               & 11:23:09.4 & +05:30:18   & 2.47  & 29.16 & OT2   &  1342256857 - 1342256861& \\
8   & 4C23.56		     & 21:07:14.8 & +23:31:45   & 2.48  & 28.93 & OT2 &  1342244144 - 1342244148& \citet{tanaka}\\
9   & MRC 2104-242         & 21:06:58.1 & -24:05:11   & 2.49   & 28.84 & OT1   &  1342218680 - 1342218684& Cooke et al. (in prep)\\
10  & MPJ 1755-6916	     & 17:55:29.8 & -69:16:54   & 2.55  & 28.95 & OT2 &  1342241081 - 1342241085&\\
11  & PKS 0529-549	     & 05:30:25.4 & -54:54:21   & 2.58  & 29.16 & OT2 &  1342241095 - 1342241099&\\
12  & NVSSJ111921-363139   & 11:19:21.8 & -36:31:39   & 2.77  & 29.39 & OT2   &  1342247263 - 1342247267&\\ 
13  & 4C+44.02             & 00:36:53.5 & +44:43:21   & 2.79   & 28.98 & OT2   &  1342249247 - 1342249251&\\           
14  & MRC 0052-241         & 00:54:29.8 & -23:51:32   & 2.86   & 28.77 & OT1   &  1342234700 - 1342234704&\citet{venemans2007}\\
15  & 4C24.28		     & 13:48:14.8 & +24:15:50   & 2.89  & 29.05 & OT2 &  1342247278 - 1342247282&\\
16  & MRC 0943-242         & 09:45:32.8 & -24:28:50   & 2.92   & 28.62 & OT1   &  1342232359 - 1342232363&\citet{venemans2007}\\
17  & MRC 0316-257         & 03:18:12.0 & -25:35:11   & 3.13    & 28.95 & OT1   &  1342237526 - 1342237530&\citet{venemans2007}\\
18  & 6CE1232+3942	     & 12:35:04.8 & +39:25:39   & 3.22  & 28.99 & OT2 &  1342247887 - 1342247891&\\
19  & 6C1909+72	           & 19:08:23.7 & +72:20:12   & 3.54  & 29.12 & OT2   &  1342241141 - 1342241145&\\
20  & 4C 1243+036          & 12:45:38.4 & +03:23:20   & 3.56   & 29.23 & OT2   &  1342259421 - 1342259425&\\
21  & MG 2141+192	     & 21:44:07.5 & +19:29:15   & 3.59  & 29.08 & OT2 &  1342245421 - 1342245425&\\
22  & TN J1049-1258        & 10:49:06.2 & -12:58:19   & 3.70  & 28.94 & OT2   &  1342256669 - 1342256673&\\
23  & 4C60.07		     & 05:12:54.8 & +60:30:51   & 3.79  & 29.20 & OT2 &  1342241135 - 1342241139&\\
24  & 4C 41.17             & 06:50:52.2 & +41:30:31   & 3.79  & 29.18 & OT2   &  1342251943 - 1342251947&\\
25  & TN J2007-1316        & 20:07:53.2 & -13:16:44   & 3.84  & 29.13 & OT2   &  1342255077 - 1342255081&\\ 
26  & PKS 1338-1942        & 13:38:26.1 & -19:42:31   & 4.11    & 28.70 & OT1   &  1342236181 - 1342236185&\citet{venemans2007}\\
\hline
\end{tabular}
\caption{\protect\label{obs_table} The radio galaxy fields targeted in the SPIRE observations, listed in order of increasing redshift, together with their $L_{\rm 500 MHz}$ radio power \citep[taken from the compendium of][]{miley2008}, the {\it Herschel} Open Time programme, OT1 or OT2, under which they were observed, and the unique observation identification numbers (OBSIDs) assigned to the data. Reference are given for the 8 fields in the sample previously identified as containing rich protoclusters.}
\end{table*}

\section{{\it Herschel} data reduction}
\label{dr}

SPIRE observes simultaneously in three far-infrared bands -- 250, 350 and 500\mic. These bands have measured full-width-half-maxima (FWHM) of approximately 18, 25 and 36\arcsec respectively. 

The observations were performed using the `large scan-map' observing mode.  The `nominal' scanning rate of 30\arcsec/s was used and two cross-linked scans were performed per map repeat.  A total of five map repeats were executed for each target field.  Following analysis of our first target field, MG 2308+0336, we implemented a dither to our observing files, wherein each repeat map is off-set from the previous one.  A dither of approximately one 250\,$\mu$m map pixel was used.  These dithered observations provide increased uniformity in observing time and redundancy across the map, and were beneficial during source extraction.

The SPIRE data were reduced using version 9.0 of the Herschel interactive pipeline environment \citep[HIPE;][]{hipe}.  The \verb1naiveScanMapper1 task within HIPE was used to make maps with a na\"ive map projection, and data obtained during the telescope scan turn-around phase were included to minimise the instrumental noise in the maps.  All other tasks, including deglitching, were used as implemented in the default SPIRE data reduction pipeline, with the nominal values.  We adopt a flux calibration error for these data of 7\% as recommended in the SPIRE Observer's Manual (SPIREOM)\footnote{$\rm http://herschel.esac.esa.int/Docs/SPIRE/html/spire\_om.html$}. The output map pixel scales are 6, 10 and 14\arcsec per pixel.

\section{Catalogue creation}
\label{cat}

This Section describes the creation of catalogues from the SPIRE maps of the HzRG fields. Two methods are used: the first considers each SPIRE band separately, whereas the second builds on the first to produce a three-band matched catalogue from which high-redshift sources can be more easily selected.

\subsection{Individual catalogues: {\small STARFINDER}}

The first source extraction method uses the {\small STARFINDER} algorithm \citep{starfinder} to create catalogues at 250, 350 and 500\mic for each HzRG field in turn. {\small STARFINDER} was designed to analyse crowded stellar fields and works by iteratively fitting a point spread function (PSF) template to potential point sources identified above some background threshold. It is therefore ideal for identifying faint sources in these confusion-limited SPIRE images. 

In practice {\small STARFINDER} is run in two stages for each band. The first pass creates a background, determined using a box size of $5 \times 5$ pixels, which is then subtracted from the original map; the extraction process is then repeated on the background-subtracted map to give the final catalogue. In both cases a two-dimensional Gaussian approximation of the SPIRE beam for the wavelength in question is used  as the template PSF. The major and minor axes FWHM used for the Gaussians are $(18.7,17.5)$\arcsec, $(25.6,24.2)$\arcsec and $(38.2,34.6)$\arcsec for the 250, 350 and 500\mic respectively; these parameters are taken from the SPIREOM. The position angle of the Gaussian in each case is set to that of the map in question. Note that this process is optimised for point sources and will underestimate the flux densities for sources larger than the SPIRE beams. This will generally only be an issue for bright, extended, nearby sources, which will not be members of the target protocluster, so can be ignored here.

The SPIRE data are calibrated in Jy/beam, so the flux density for a particular source comes directly from the peak of the Gaussian fit. However, the pixelisation of the time-line data suppresses the signal in a map pixel, which means the peak fit will underestimate the true value. Correction factors for 250, 350 and 500\mic of 4.9, 6.9 and 9.8 per cent (again taken from the SPIREOM) are needed to account for this, and have been incorporated into the final catalogues. 

The flux density errors determined by {\small STARFINDER} are combined in quadrature with the confusion noise in each band \citep[5.8, 6.3 and 6.8 mJy/beam at 250, 350 and 500\mic;][]{nguyen} to give the final 1$\sigma$ uncertainties. Taking this into account, the average 3$\sigma$ flux density limits of the catalogues are 21.3, 24.6 and 25.5 mJy/beam for 250, 350 and 500\mic respectively. These limits correspond to an infrared luminosity of $\sim 10^{12.5}$ L$_{\subsun}$ at $z=2$, increasing to $\gtrsim 10^{13}$ L$_{\subsun}$ at $z=4$. A further conversion to star--formation rate, assuming a Salpeter initial mass function, \citep{kennicutt1998a,kennicutt1998b} gives detection limits for these data of $\sim 500$ and $\gtrsim 1700$ M$_{\subsun}$/yr for redshifts of 2 and 4 respectively. 

\begin{figure*}
\centering
\vspace{-5mm}
\includegraphics[angle=-90, scale=0.5]{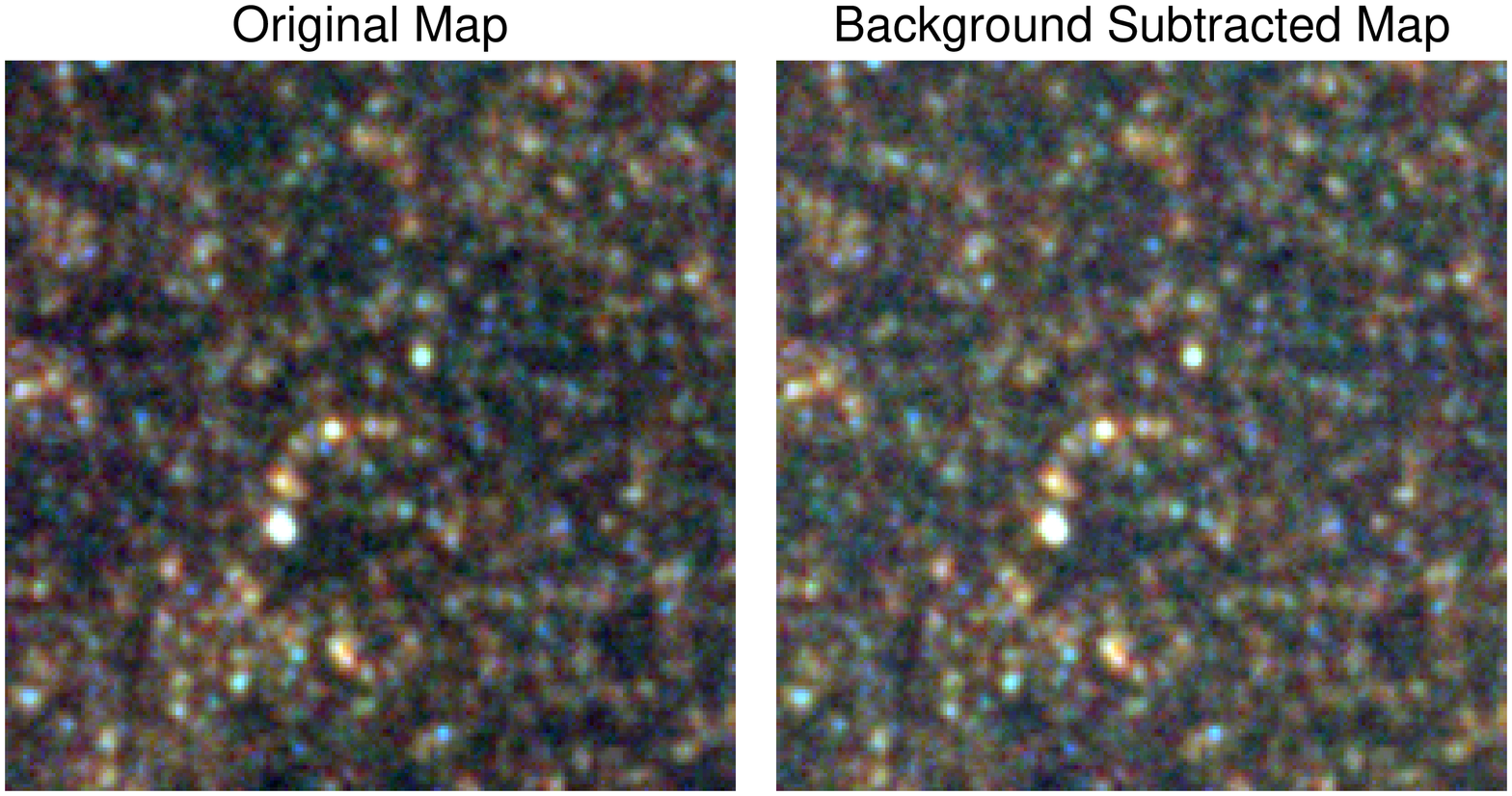}
\vspace{-7mm}
\caption{\protect\label{no_cirrus} False colour, combined 250, 350 and 500\mic, images of the central 20\arcmin $\times$ 20\arcmin region of the MRC 1138-262 field; both are displayed using the same colour scaling. 250, 350 and 500\mic data are represented by blue, green and red colours respectively. This field is not strongly affected by background contamination (left), which the background--subtraction successfully reduces (right).  }
\vspace{-5mm}
\includegraphics[angle=-90, scale=0.5]{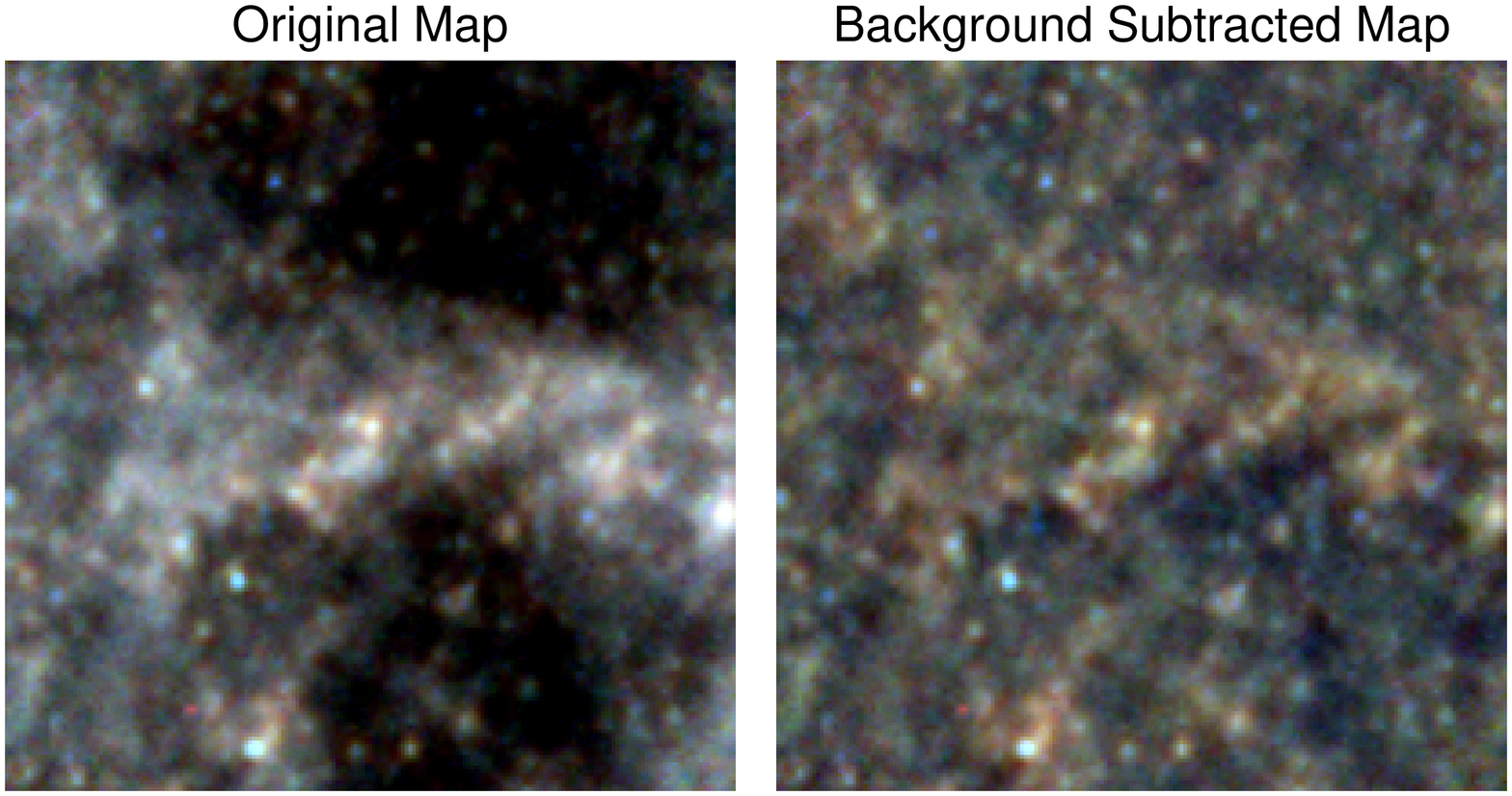}
\vspace{-7mm}
\caption{\protect\label{bad_cirrus} False colour, combined 250, 350 and 500\mic, images of the central 20\arcmin $\times$ 20\arcmin region of the 4C60.07 field; both are displayed using the same colour scaling. 250, 350 and 500\mic data are represented by blue, green and red colours respectively. The image on the left is before background subtraction and shows clear contamination, mainly by foreground galactic cirrus. The one on the right is background--subtracted; the contamination is reduced but some underlying cirrus structure remains.}
\end{figure*}

\subsubsection{The effect of Galactic cirrus}

A major source of contamination in extragalactic far--infrared maps is foreground cirrus emission from our own Galaxy. This peaks at $\sim$200\mic \citep{bracco}, so is brightest in the shortest SPIRE band, but it can also be an issue at the longer wavelengths because of their lower resolution. The main effect of this cirrus is to increase the confusion in the maps, but the small--scale structure within it can also lead to spurious detections in the catalogues. 

Inspection of the outputs from {\small STARFINDER} shows that in the majority of cases, the background--subtraction procedure is sufficient to account for this contamination (Figure \ref{no_cirrus}). However, cirrus structure remains in a small number of the worst--affected fields even after the background is removed (Figure \ref{bad_cirrus}). A cirrus--free field would be expected to have a flat power spectrum at spatial frequencies larger than the instrument beam, in contrast to a field with high galactic cirrus confusion, which would exhibit a negative power spectrum on these scales at SPIRE wavelengths \citep{martin2010}. The spectral index of the field power spectrum, $\alpha$, can therefore be used as a metric to identify high cirrus fields. In practice, however, this flat spectrum assumption may not be valid because of the presence of large scale structure in the SPIRE data. Consequently a high--cirrus field here is identified by comparing the $\alpha$ of each original 250\mic map (chosen because the cirrus signal should be strongest there) with the spectral index, $\alpha_{\rm ref}$, of a control field which is known to be cirrus--free (see Section \ref{ref_field_sect} for a full description of this reference). Affected fields are taken to be those with spectral index steeper than $\alpha < (\alpha_{\rm ref} - 0.25)$; the success of this classification is confirmed visually. 

There are 7 fields flagged by this process: 3C257, 4C23.56, PKS 0529-549, 6C1909+72, MG 2141+192, 4C60.07 and 4C41.17. Of these, only 4C23.56, is a previously known protocluster field. They are excluded from the subsequent overdensity analysis presented here.  

\subsection{Three-band matched catalogues}

The catalogues described above are designed to examine the behaviour of the source population in each far--infrared band separately. However, it is also useful to be able to compare the flux densities of sources between bands. A high--redshift source, for example, will generally be brighter at 500\mic than at 250\mic \citep[e.g.][]{amblard}. A simple cross--matching between the catalogues is insufficient for this, given the decreasing resolution, and consequent higher level of source blending, at the longer wavelengths which may bias sources towards these redder far--infrared colours. The solution to this problem, described in detail below, is a combined catalogue created using information from all three bands. 

It should be noted that whilst this method uses the shorter SPIRE wavelengths to overcome some of the blending issues at 500\mic, it cannot account for sources which are blended at 250\mic (where the beamsize is $\sim$18\arcsec). This problem is reduced by limiting the analysis to $\geq 3\sigma$ detections only, and, since the overdensity analysis used to identify potential protoclusters is carried out via comparison to a reference field (see Section \ref{env} for more details), the overall effect of this on the results is negligible.

\subsubsection{250\mic prior catalogues}

Synthetic maps of 350\mic and 500\mic sources are created based on the positions of the 250\mic sources from the {\small STARFINDER} catalogue; this is limited to $>3\sigma$ sources only. A map of each source is created using a Gaussian with a FWHM matched to the {\it Herschel} beam size at the appropriate wavelength. The fluxes of each source are determined one-by-one starting with the brightest 250\mic source. A combined map of all the sources within a box of 84\arcsec\ length centred on the 250\mic source is created. The flux of each source within this box is allowed to vary using the IDL function {\small AMOEBA}, until the combined synthetic map matches the observed map. This allows us to take into account source blending in the 350\mic and 500\mic images. Fits to the 350\mic and 500\mic maps are performed separately and the flux ratios are allowed to vary freely as no template SED is assumed.

\subsubsection{500\mic prior catalogues}

The matched catalogue described above is potentially biased against high redshift sources, which are typically brightest at 500\mic. To compensate for this a new source catalogue is created based on the 500\mic catalogue; this is again limited to $>3\sigma$ sources only. Counterparts for each 500\mic source are sought in the higher-resolution and deeper 250\mic and 350\mic images. We first search for a counterpart in the 250\mic catalogue within 10$\arcsec$ of the 500\mic position. If a counterpart, or multiple counterparts, are detected the positions of these counterparts are entered into the catalogue. If no counterpart is found in the  250\mic catalogue, the search is repeated  in the 350\mic catalogue, again within 10$\arcsec$ of the 500\mic position. The positions of any detected  350\mic counterparts are entered into the new combined catalogue, but if no counterpart is found in either catalogue, we enter the 500\mic position into our combined catalogue.

Synthetic maps of 250\mic, 350\mic and 500\mic sources are created based on the positions of this new 500\mic prior catalogue. Each source is modelled by a Gaussian with a FWHM matched to the {\it Herschel} beam size at the appropriate wavelength. In order to take into account blending, and the flux from nearby sources, a map is created of all sources from the 500\mic prior catalogue that lie within a box of 84\arcsec on a side, centred on the source being measured. The flux density of each source within this box is allowed to vary using the IDL function {\small AMOEBA}, until the 84\arcsec-sized synthetic map matches the observed map. The flux density of the central source in this box is then fixed to the best-fit value before moving on to measure the next brightest source. Fits to the 250\mic, 350\mic and 500\mic maps are performed separately and the flux ratios are allowed to vary freely as no template SED is assumed.

In the remainder of this paper MC250 and MC500 will be used to refer to the 250\mic and 500\mic prior three--band matched catalogues respectively. 

\subsection{Catalogue completeness}

The completeness of the individual {\small STARFINDER} catalogues is estimated by inserting fake sources into the maps and extracting them as described above. These are simulated as Gaussian point sources with dimensions equal to that of the instrument beam at the wavelength in question. As the images are confusion-limited the density of sources is high. Therefore, to minimise blending with real galaxies, only 20, all set to the same flux density, are introduced each time. After 10 repetitions, the fake flux density is increased and the process restarted, giving a final total of  200 fake sources per field per band per flux density.  This global determination of completeness is appropriate here as the observations are all to the same depth.

The 80 per cent completeness level is found to be 28 mJy at 250\mic, 36 mJy at 350\mic, and 36 mJy at 500\mic, and the completeness at the 3$\sigma$ flux density limit in each band is 62\%, 62\% and 61\% at 250, 350 and 500\mic respectively. The equivalent completeness in the MC250 and MC500 catalogues can be taken as the completeness in the 250 and 500\mic catalogues upon which they are based. 

\section{Searching for overdensities in the HzRG fields}
\label{env}

\begin{figure*}
\centering
\includegraphics[scale=0.35, angle=-90]{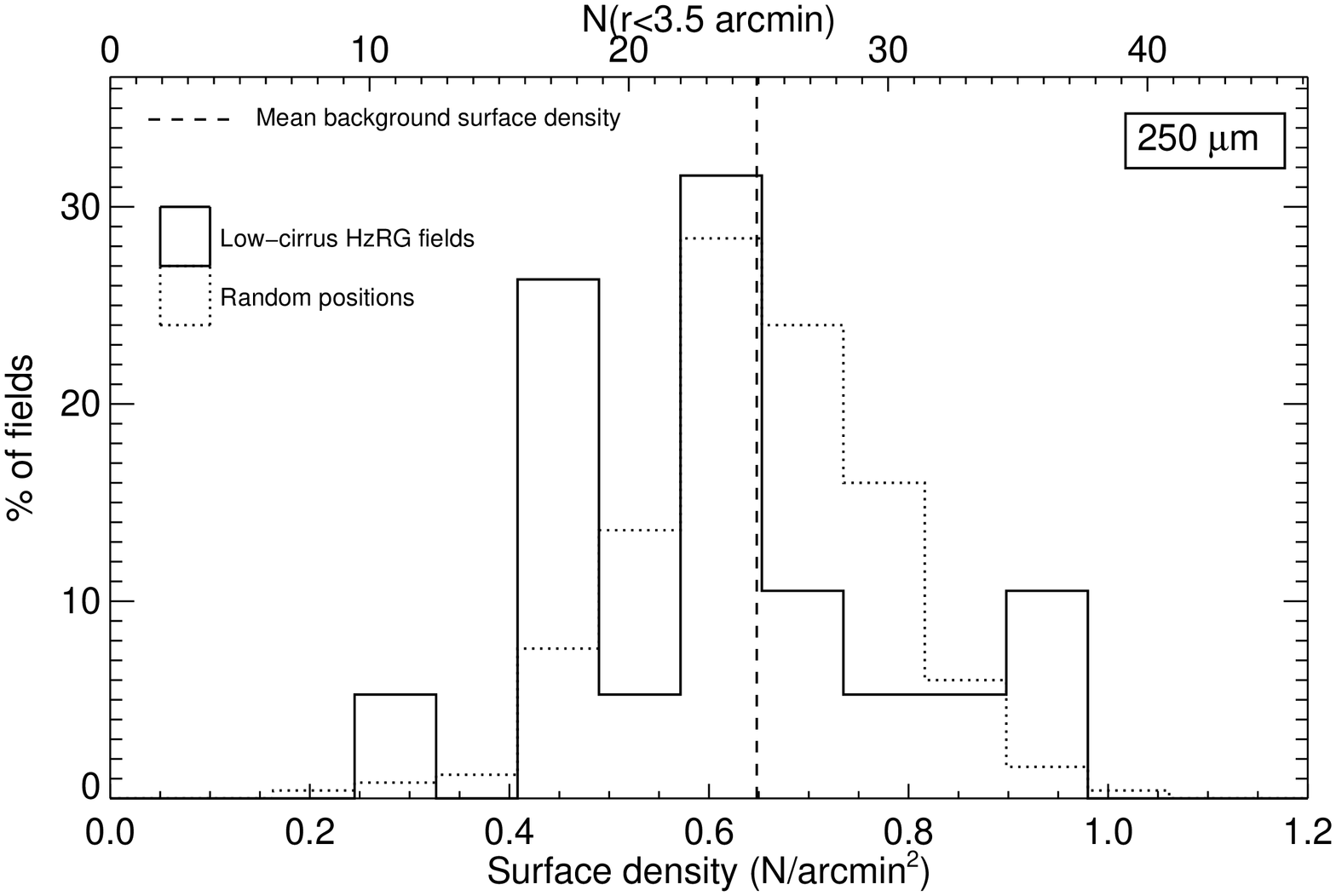}
\includegraphics[scale=0.35, angle=-90]{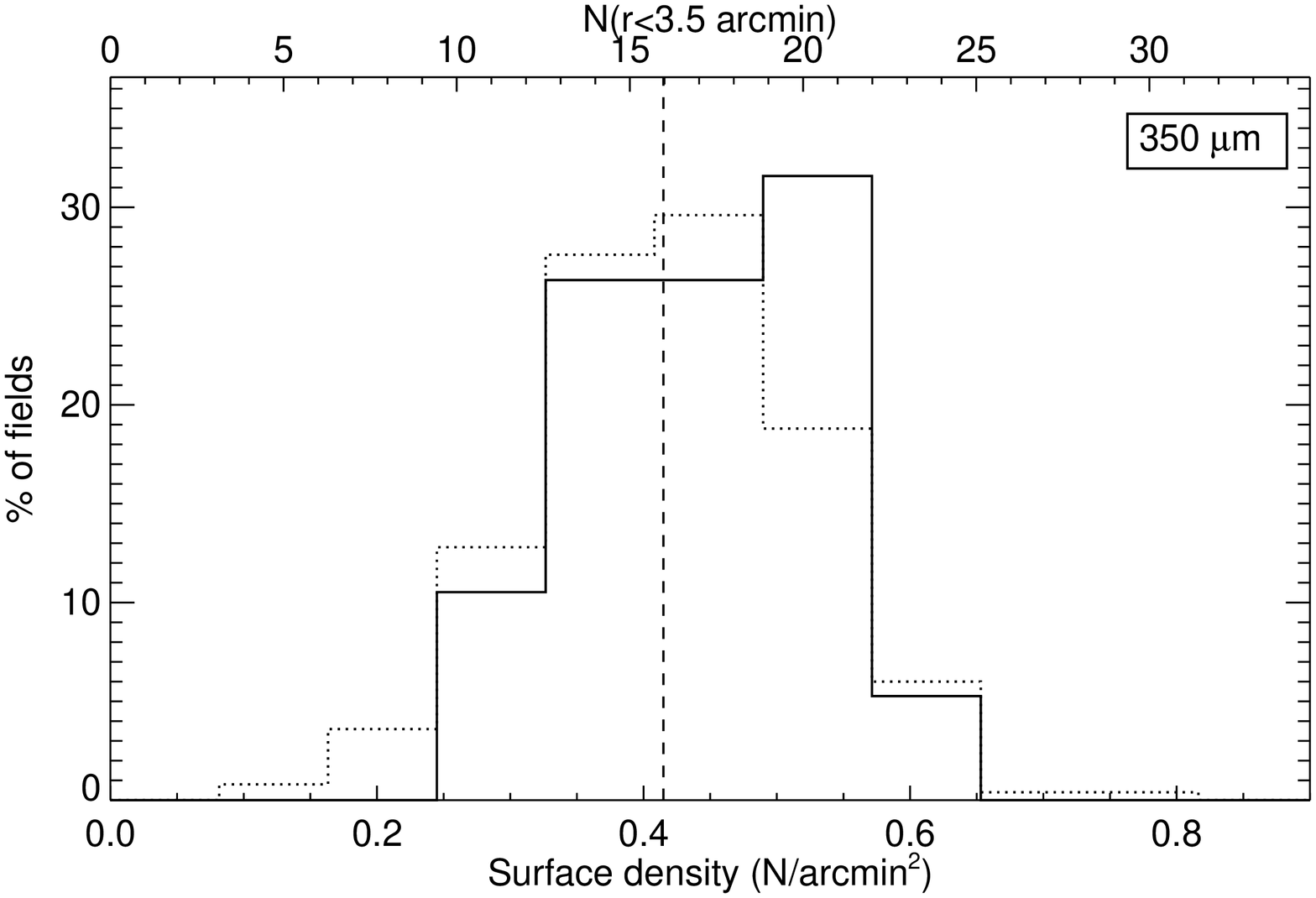}
\includegraphics[scale=0.35, angle=-90]{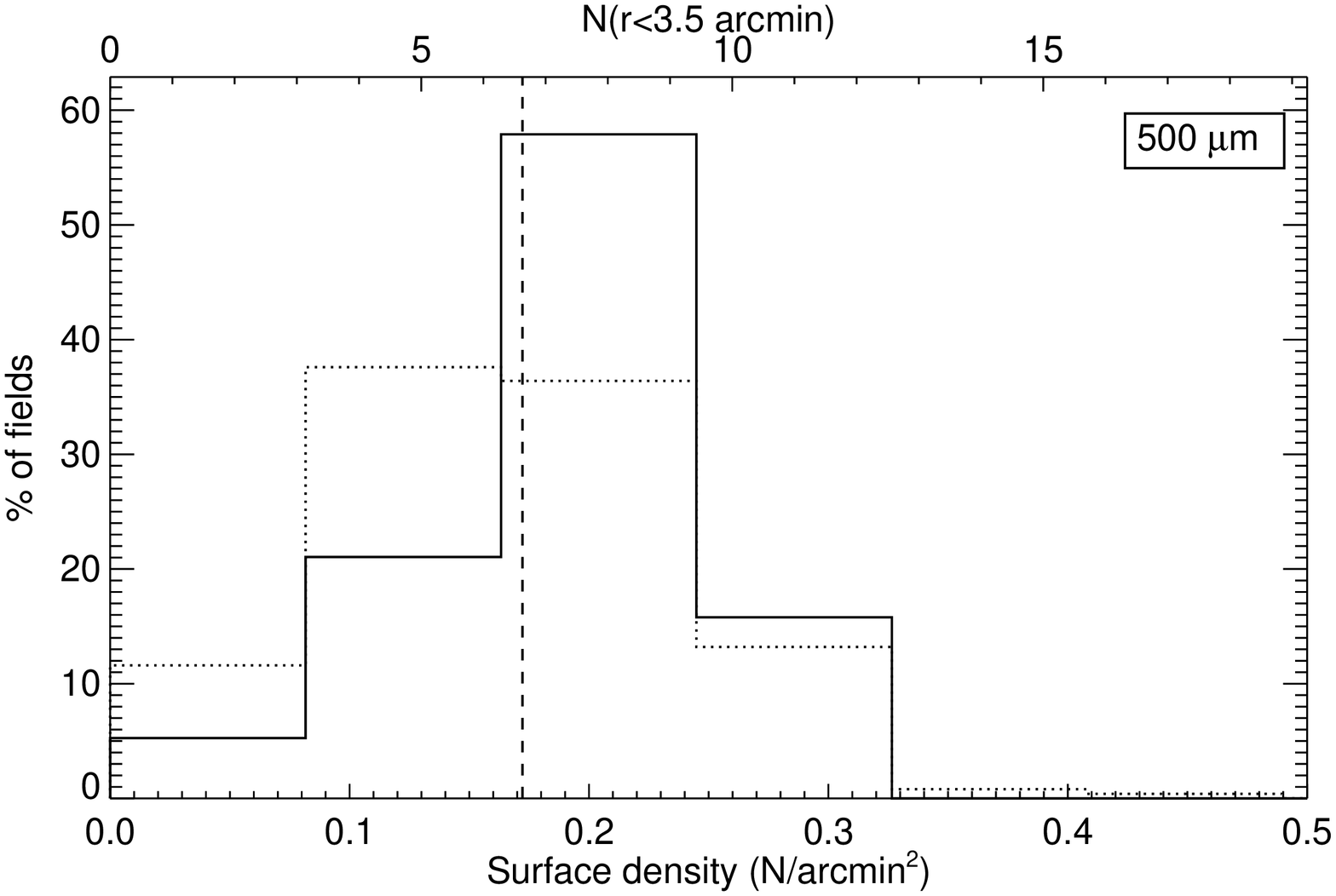}
\caption{\protect\label{sd_indiv} The distribution of surface density within a radius of 3.5\arcmin from the central HzRG position, calculated using the individual band catalogues. The dotted histogram shows the results for 250 random positions in the reference field; the mean of this distribution is indicated by a vertical dashed line. The solid histogram shows the surface densities for the 19 low--cirrus radio galaxy fields.}
\end{figure*}

\subsection{The reference field}
\label{ref_field_sect}

The identification of overdensities in the HzRG fields is only possible if there is a reference field with which to compare the results. This should ideally have been observed using the same observing mode and to the same depth as the HzRG maps. These criteria are satisfied with a custom reference field which is created by reducing the first five map repeats of a 2.8 sq. deg. observation of the COSMOS field (centred at $\alpha$=10$^{h}$00$^{m}$28.6$^{s}$, $\delta=+$2\deg12\arcmin21\arcsec [J2000]) using the same method described in Section \ref{dr}. These data are drawn from the {\it Herschel} Science Archive and were originally observed as part of the {\it Herschel} Multi--tiered Extragalactic Survey \citep[HerMES][]{glenn, oliver}. This field contains an overdensity of sub--millimetre galaxies at $z \sim 1$ \citep{austermann}, but no known overdensities at $z>2$ which could affect its effectiveness as a blank reference area for this 

Sources are extracted from the custom HerMES maps using the same methods as those used for the HzRG field to create reference versions of the individual and matched catalogues.

\subsection{Surface density analysis}

The potential protocluster environment around the HzRG in the {\it Herschel} fields can be investigated using the source surface density i.e. the number of sources within some radius. \citet{hatch2011a} found that the protocluster galaxies in their $z \sim 2.5$, near--infrared, sample lie within 3.5\arcmin of the HzRG. This, therefore is adopted as the initial search radius here and corresponds to $\sim$6.5 comoving Mpc at $z = 3$.

Figure \ref{sd_indiv} shows the results of the surface density calculation, using the three individual catalogues, each cut to include $\geq3\sigma$ sources only. 
Also shown is the background surface density, determined by counting sources in the reference catalogue that lie within a 3.5\arcmin radius of 250 random positions in the reference field. It should be noted that the central HzRG is only detected in $\sim$50\% of the fields at 250\mic. The AGN typically contributes little to the far--infrared flux density and may quench the star formation in its host galaxy. Indeed, radio--detected galaxies have been found to be typically fainter than their non radio--detected counterparts \citep{virdee}. 

It is clear from Figure \ref{sd_indiv} that the proportion of low-cirrus fields with densities higher than the mean background level increases with wavelength. At 250\mic the distributions of the random and HzRG fields are similar, but at 500\mic the majority of HzRG fields are more dense than the background within the search region. This is not unexpected as higher redshift sources typically have redder far--infrared colours than lower redshift ones, due to the rest--frame SED dust peak moving through the SPIRE bands. A two--sided Kolmogorov--Smirnov test gives a probability of 10\% that the HzRG fields are consistent with the background at 500\mic, and thus suggests that this overdensity excess is only marginally significant. 

\begin{figure*}
\centering
\includegraphics[scale=0.35, angle=-90]{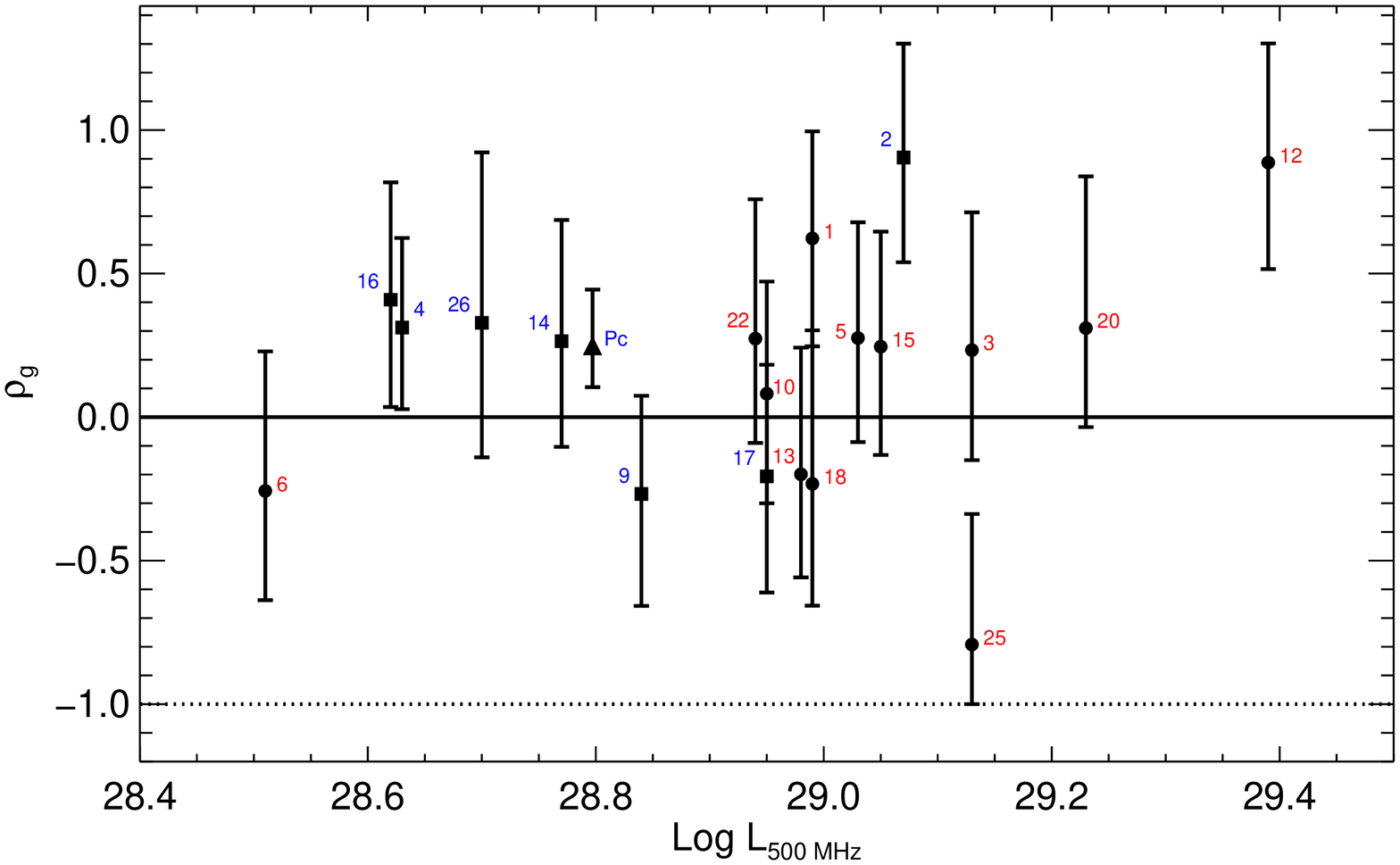}
\includegraphics[scale=0.35, angle=-90]{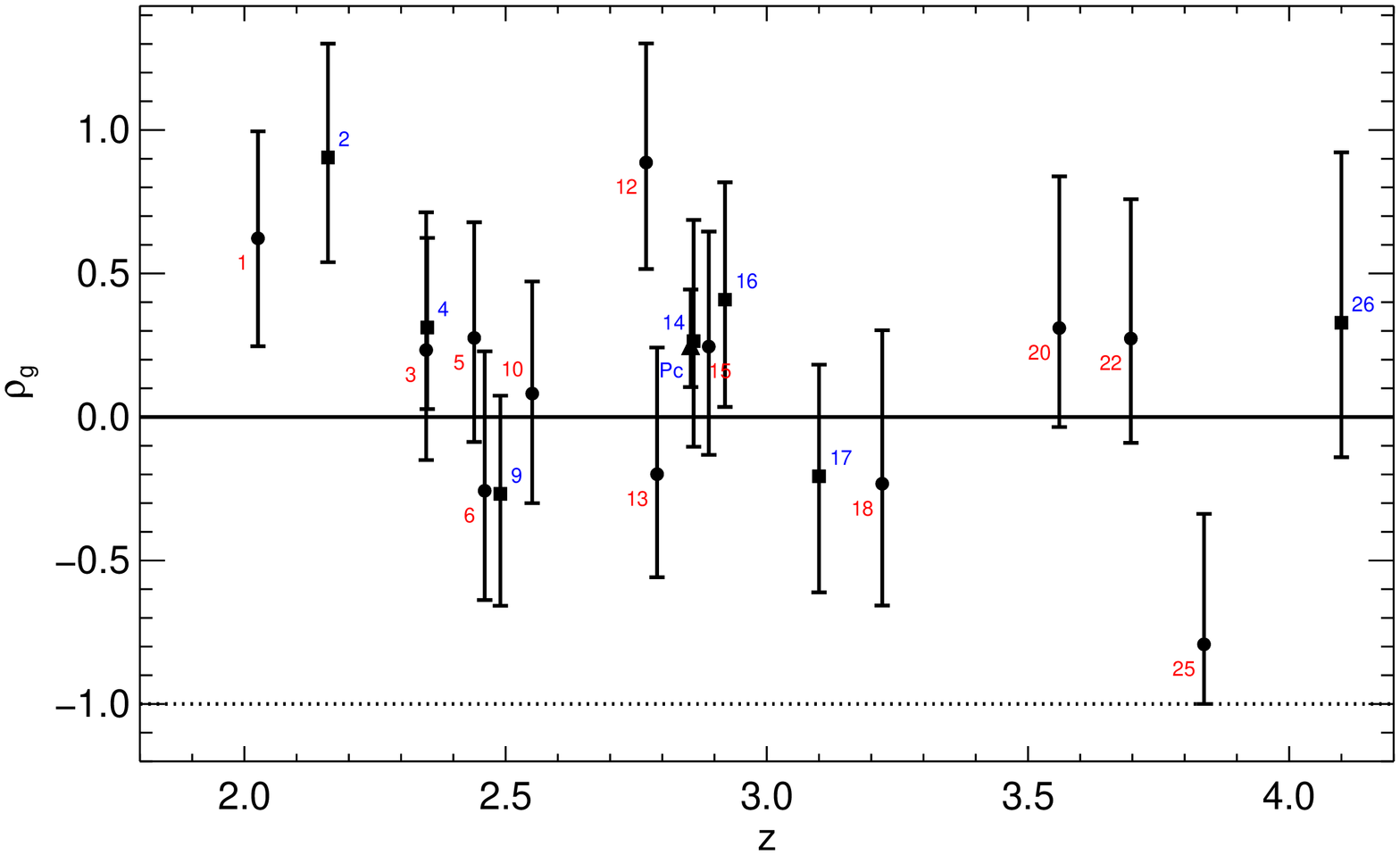}
\caption{\protect\label{excess_500} The overdensity of galaxies in each field as a function of radio power (left) and redshift (right) of the central HzRG.  Galaxies are selected from the 500\mic catalogue and counted within a radius of 3.5\arcmin of the position of the HzRG; only the 19 low--cirrus fields are considered. Numbers correspond to the field labels given in Table \protect\ref{obs_table}. Blue colours and square symbols highlight the 7 known protoclusters, and the average overdensity of these fields is labelled `Pc'. A value of $\rho_{\rm g} = -1$ (indicated by the horizontal dotted line) corresponds to a field with no galaxies within the search radius.}
\end{figure*}

\subsection{Quantifying the galaxy overdensities}

The overdensity of galaxies in the fields is defined as the excess surface density, $\rho_{g}$, calculated by
\begin{equation}
\label{eqn}
\rho_{g} = (\rho_{\rm obs} - \rho_{\rm bkg}) / \rho_{\rm bkg} , 
\end{equation}
where $\rho_{\rm obs}$ and $\rho_{\rm bkg}$ are the observed surface density and mean background surface density respectively. The bias in the underlying galaxy population means that the variation in background surface density across the random search regions deviates from the expected Poisson distribution. The upper and lower uncertainty bounds on $\rho_{g}$ are therefore taken as the 16th and 84th percentiles of the background surface density distribution. 

This technique is particularly useful as it allows custom search radii to be used for each field. As a result the search radius here is updated from a constant map scale (3.5\arcmin) to a constant comoving scale.  Inspection of the data the shows that the highest significance of overdensity for the largest number of fields occurs at comoving radii of between 6 and 7 comoving Mpc.  No significant variation was found when using radii between 6 and 7 comoving Mpc, therefore a scale of 6 comoving Mpc is adopted for the remainder of this work.  This radius equates to angular scales ranging from 4\arcmin at $z=2$ to 2.8\arcmin at $z=4$. Keeping this value constant across the 19 fields allows their results to be easily compared, though any overdensities present may in principle be different sizes in different fields. 

Applying Equation \ref{eqn} to the 500\mic catalogues, where the overdensities are strongest, gives the results shown in Figure \ref{excess_500}, plotted against both redshift and radio luminosity of the HzRG. The corresponding overdensity values are given in Table \ref{dens_table}, together with the number of galaxies found in each search region. Only the 19 low--cirrus fields are included here. The trend towards overdensity in the fields seen in Figure \ref{sd_indiv} is replicated here, but the 5 fields with a lower limit on $\rho_{g} > 0$ (MPJ 1758-6737, MRC 1138-262, USS 1707+105, NVSSJ111921-363139 and MRC 0943-242) are not significantly overdense ($\leq 2.6\sigma$). Of the 8 previously known protoclusters (labelled `Pc' in Table \ref{dens_table} and highlighted in blue in Figure \ref{excess_500}) only 3 show this behaviour, with the average `Pc' overdensity consistent with the background. Finally, there is no significant correlation with either redshift or radio luminosity of the HzRG. 

\subsubsection{HzRG redshift colour selection}

An improved determination of the far--infrared galaxy overdensities present at 500\mic in the potential protocluster fields is possible by selecting the sources which lie near the HzRG in redshift space. A simple colour cut ($f_{350 \mu m}/f_{250 \mu m} \geq 0.85$ for instance) will select sources with $z \gtrsim 2$ \citep[e.g.][]{amblard}, but, since the HzRGs cover $2 < z \lesssim 4$, this could result in contaminating foreground galaxies remaining in a large proportion of the fields.  A more sophisticated approach is to use a custom--colour selection, tailored to the redshift of each HzRG. 

\begin{figure}
\centering
\includegraphics[angle=-90, scale=0.35]{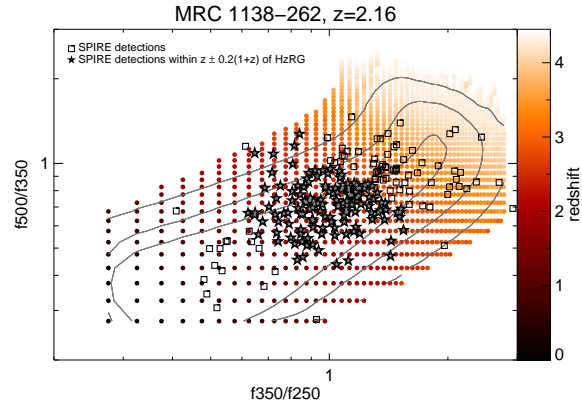}
\caption{\protect\label{col_plot} SPIRE colour--colour diagram for one of the HzRG fields, illustrating the custom--colour redshift selection. Sources in the MC500 catalogue are selected to lie within $\pm 0.2(1+z)$ of the redshift of the HzRG by comparing their SPIRE colours with those of an artificially redshifted galaxy template (constructed following Roseboom et al. 2012). Stars and squares indicate sources selected and rejected as potential protocluster members respectively; solid points show the colour track of the template galaxy, broadened by a Gaussian uncertainty of 10\%, as it evolves in redshift. Lines show the 1, 2 and 3$\sigma$ contours in the template distribution.}
\end{figure}

Figure \ref{col_plot} illustrates this technique; sources in a matched catalogue are selected to lie within some $\delta z$ by comparing their SPIRE colours with those derived from an artificially redshifted modified black body dusty galaxy SED template, with a temperature of $T = 40$ K, and a dust emissivity parameter of $\beta = 1.8$. The parameters are taken from \citet{roseboom}, who found that this gave a reasonable match to their far--infrared sources. A Gaussian standard deviation of 10\% is also added to the generated template flux densities to account for the spread in the real data; as Figure \ref{col_plot} shows, the number of sources which lie outside the limits defined by this model is low. The redshift selection range is taken as $\pm 0.2(1+z)$; this is the uncertainty expected for a `reliable' photometric far--infrared redshift in \citeauthor{roseboom}. There is a degeneracy between redshift and the parameters used to define the SED template which will introduce further uncertainties into this process. However, since the same redshift estimation is applied to both the reference and the HzRG fields this effect can be ignored here. 

\begin{figure*}
\includegraphics[scale=0.35, angle=-90]{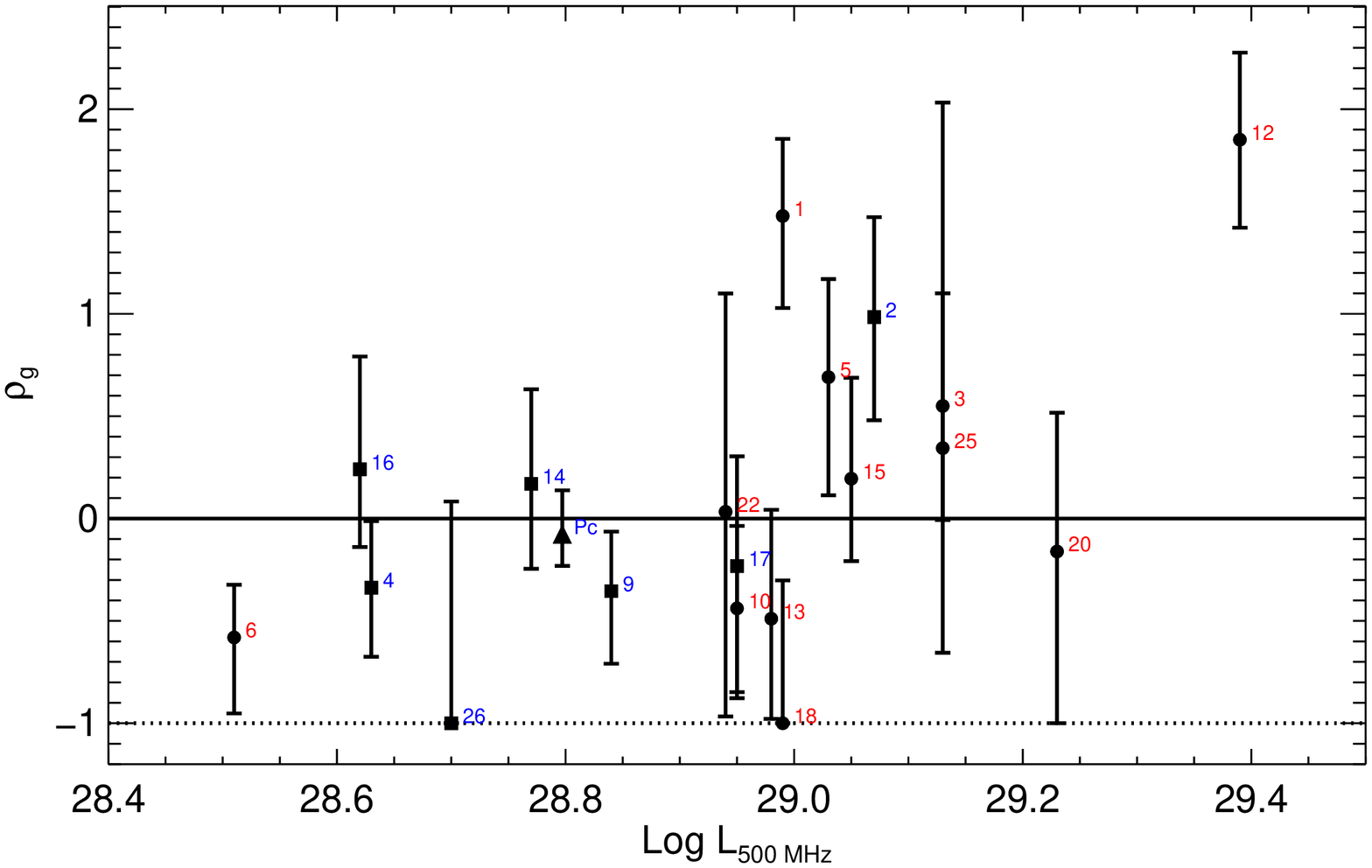}
\includegraphics[scale=0.35, angle=-90]{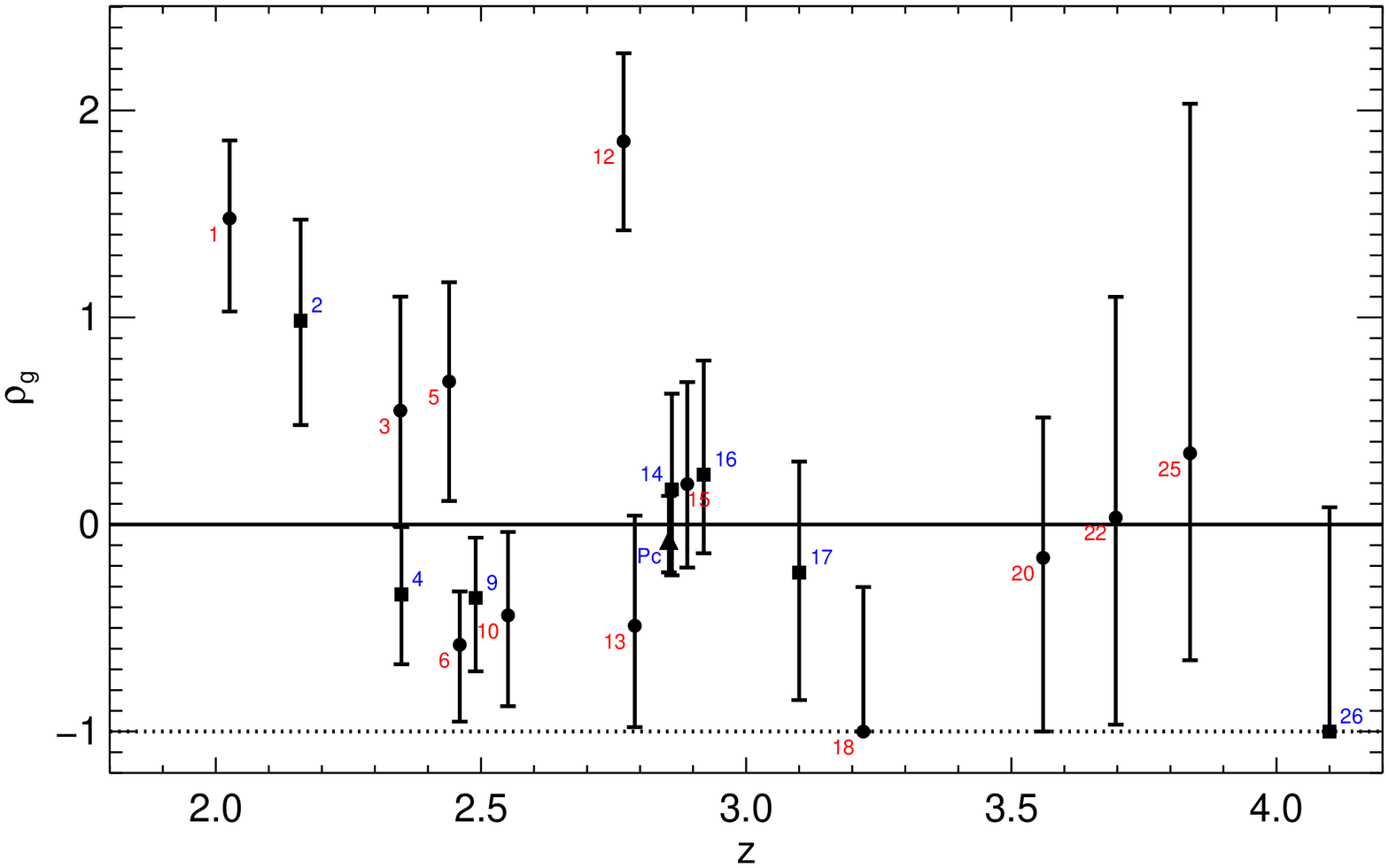}
\caption{\protect\label{excess_cc_500prior} The overdensity of galaxies in each field as a function of radio power (left) and redshift (right) of the central HzRG.  Galaxies are selected from the MC500 catalogue and counted within a radius of 6 comoving Mpc and $\pm 0.2(1+z)$ of the position and redshift of the HzRG; only the 19 low--cirrus fields are considered. Numbers correspond to the field labels given in Table \protect\ref{obs_table}. Blue colours and square symbols highlight the 7 known protoclusters, and the average overdensity of these fields is labelled `Pc'. A value of $\rho_{\rm g} = -1$ (indicated by the horizontal dotted line) corresponds to a field containing no galaxies which satisfied the position and redshift criteria.
}
\end{figure*}

This colour selection is applied to the low--cirrus three--band matched catalogues (MC500) and the galaxy overdensities within 6 comoving Mpc are calculated as before. The background galaxy density is determined separately for each field by applying the same colour selection to the appropriate MC500 version of the reference catalogue, and then searching for the sources within it that lie around 250 random positions in the reference field.

Figure \ref{excess_cc_500prior} shows the results, again plotted against the redshift and radio luminosity of the HzRGs. The corresponding overdensity values are given in Table \ref{dens_table}. In contrast to the results obtained using the individual 500\mic catalogue, 2 fields are $> 3\sigma$ overdense here. No fields display underdensities of a similar large significance. Only one of the 8 known protocluster fields, MRC 1138-262, has $\rho_{g} > 0$, but this overdensity is only $2\sigma$.
Visual inspection of the Figure reveals that the larger overdensities tend to lie at $L_{\rm 500 MHz} \gtrsim 10^{29}$ and $z < 3$, but the Spearman rank correlation co--efficients show that there are no significant correlations with overdensity for either of these parameters. However, this apparent preference for brighter HzRGs is consistent with previous work \citet[][and references therein]{galametz}. 

The spatial distribution of the colour--selected sources in each field is shown in Appendix \ref{ra_app}. The central HzRG is only detected and colour--selected in 26\% of the fields (rising to 50\% for fields with $\rho_{\rm g} > 0$). 

At the highest redshifts probed the colour--selected source density is low, and the search radius contains at most 1 source in all but one of the fields. Inspection of the spatial distribution of sources for these fields (Appendix \ref{ra_app}) shows that typically 1 -- 5 potential protocluster members are present, but located at 5\arcmin to 10\arcmin ($\sim$10 -- 20 comoving Mpc at $z \sim 3.5$) radii from the HzRG position. One possible explanation for this is that the most distant protoclusters are larger than their lower redshift counterparts. However, simulations predict that the majority of member galaxies should lie within 10 comoving Mpc so it is likely that this is a consequence of the large uncertainty in the redshift selection.

The colour--selected overdensities can also be calculated using the MC250 catalogues as a consistency check. This shows a similar overall behaviour in $\rho_{g}$, and the 2 fields with the strongest MC500 overdensities are also significantly overdense here ($\rho_{g} = 1.25^{+0.35}_{-0.40}$ and $1.81^{+0.53}_{-0.49}$ for MP J1758-6737 and NVSS J111921-363139 respectively).

\begin{table}
\centering
\begin{tabular}{lllll}
\hline
&      &      & 500\mic      & MC500         \\
\multicolumn{2}{l}{Label}  & Name & $\rho_{\rm g}$ & $\rho_{\rm g}$ \\
\hline
1 &  & MP J1758-6738        &      0.62$^{+0.37}_{-0.38}$  (13)  &             {\bf 1.48$^{+0.38}_{-0.45}$} (9)  \\                                 
2 & Pc & MRC 1138-262       &      0.90$^{+0.40}_{-0.37}$  (15)  &                 0.98$^{+0.49}_{-0.50}$  (8)  \\          
3  & & BRL 0128-264         &      0.23$^{+0.48}_{-0.38}$  (10)  &                 0.55$^{+0.55}_{-0.56}$  (7)  \\          
4  &Pc & USS 1707+105       &      0.31$^{+0.31}_{-0.28}$  (11)  &                 -0.34$^{+0.33}_{-0.34}$  (3)  \\         
5  & & MRC 0406-244         &      0.28$^{+0.40}_{-0.36}$  (10)  &                 0.69$^{+0.48}_{-0.58}$  (8)  \\          
6  & & MG 2308+0336         &      -0.26$^{+0.49}_{-0.38}$  (6)  &                 -0.58$^{+0.26}_{-0.37}$  (2)  \\         
9  &Pc & MRC 2104-242       &      -0.27$^{+0.34}_{-0.39}$  (6)  &                 -0.35$^{+0.29}_{-0.35}$  (3)  \\         
10 & & MP J1755-6916        &      0.08$^{+0.39}_{-0.38}$  (7)  &                  -0.44$^{+0.40}_{-0.44}$  (2)  \\         
12  && NVSS J111921-363139  &      0.89$^{+0.42}_{-0.37}$  (12)  &             {\bf 1.85$^{+0.43}_{-0.43}$} (10)  \\        
13  && 4C+44.02             &      -0.20$^{+0.44}_{-0.36}$  (5)  &                 -0.49$^{+0.53}_{-0.49}$  (2)  \\         
14  &Pc& MRC 0052-241       &      0.26$^{+0.42}_{-0.37}$  (8)  &                  0.17$^{+0.46}_{-0.42}$  (4)  \\          
15  && 4C24.28              &      0.25$^{+0.40}_{-0.38}$  (8)  &                  0.19$^{+0.49}_{-0.40}$  (4)  \\          
16  &Pc& MRC 0943-242       &      0.41$^{+0.41}_{-0.37}$  (9)  &                  0.24$^{+0.55}_{-0.38}$  (4)  \\          
17  &Pc& MRC 0316-257       &      -0.21$^{+0.39}_{-0.40}$  (4)  &                 -0.23$^{+0.54}_{-0.62}$  (2)  \\         
18  && 6CE1232+3942         &      -0.23$^{+0.53}_{-0.42}$  (4)  &                 -1.00$^{+0.70}_{-0.00}$  (0)  \\         
20  && 4C1243+036           &      0.31$^{+0.53}_{-0.34}$  (6)  &                  -0.16$^{+0.68}_{-0.84}$  (1)  \\         
22  && TN J1049-1258        &      0.27$^{+0.49}_{-0.36}$  (6)  &                  0.03$^{+1.07}_{-1.00}$  (1)  \\          
25  &     & TN J2007-1316   &      -0.79$^{+0.45}_{-0.21}$  (1)  &                 0.34$^{+1.69}_{-1.00}$  (1)  \\          
26  &Pc& PKS 1338-1942      &      0.33$^{+1.00}_{-0.47}$  (5)  &                  -1.00$^{+1.00}_{-0.00}$  (0)  \\         
\hline
     \multicolumn{3}{l}{Average of known protoclusters}    & 0.25$^{+0.20}_{-0.14}$ & -0.08$^{+0.21}_{-0.16}$ \\
\hline
\end{tabular}
\caption{The overdensity, $\rho_{\rm g}$, in each of the 19 low--cirrus protocluster field, measured within a search radius of 6 comoving Mpc, centred on the HzRG. The corresponding number of galaxies found within the search region is given in parentheses. Both catalogues include only sources which are $>3\sigma$; the MC500 catalogues are also cut to include sources lying within $\pm 0.2(1+z)$ of the redshift of the HzRG only, as implied by the far--infrared colours. The uncertainties on the measurement are taken as the 16th and 84th percentiles in the relevant background distribution. Fields with overdensities lying $\geq 3\sigma$ from the mean are highlighted in bold. The 7 fields containing known protoclusters are labelled `Pc', and their average overdensity is given in the last row. \protect\label{dens_table}
}
\end{table}

\subsection{The significance of the galaxy overdensities}

The two fields with the strongest overdensities are detected at the 3.9 and 4.3$\sigma$ level respectively (Table \ref{dens_table}). The overall significance of these results is the probability of finding 2 $\geq 3\sigma$ overdensities such as these by chance, given that 19 fields were observed. To calculate it, the probability density function for each field is randomly sampled 10,000 times, and the frequency of overdensities as strong as this is counted in each iteration. This gives a final probability of $5 \times 10^{-4}$ and suggests that the results are inconsistent with being due to random background fluctuations. 

\subsection{The radial extent of the galaxy overdensities}
\label{rad_extent_sec}

The two fields that show a significant overdensity in the MC500 catalogues can be used to measure the radial extent of the average far--infrared protocluster candidate.  This is done by counting the cumulative number of galaxies within increasing comoving radii from the HzRG positions, and subtracting the expected number of galaxies (measured from the reference field). The radial extent is the point at which the variation in this galaxy excess with radius flattens.

\begin{figure}
\centering
\includegraphics[scale=0.35, angle=-90]{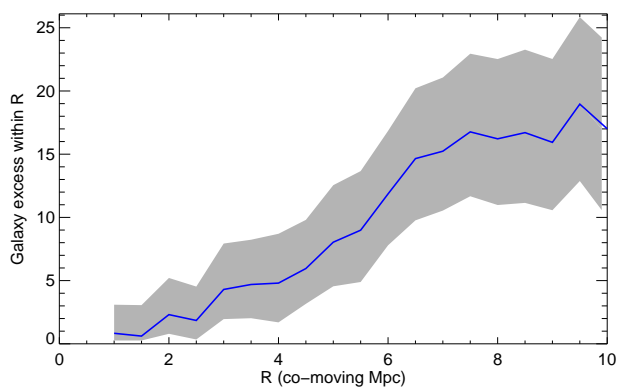}
\caption{\protect\label{rad_extent} The cumulative excess number of galaxies within a comoving radius, R, of the HzRGs in the 2 fields identified as overdense using the MC500 catalogues (MP J1758-6738 and  NVSS J111921-363139). The radial extent of the galaxy overdensity is the point at which the cumulative distribution flattens. Note that since the 2 fields are combined here, the average overdensity contains half of the numbers shown. The grey shaded area shows the uncertainty on this measurement derived from the reference catalogue and the excess number of galaxies in each bin.}
\end{figure}

This calculation shows that the average galaxy overdensity here is contained within $\sim$6 comoving Mpc, with an excess of $\sim 8 \pm 2$ galaxies (Figure \ref{rad_extent}).  This is in agreement with the radial extent determined by \citet{hatch2011a} for near--infrared selected overdensities. They also find that the majority of the excess galaxies are found in the central region ($<3$ comoving Mpc). This is not replicated here as only $\sim 1/3$ lie within this radius.

\section{Discussion}
\label{discuss}

\subsection{Are the overdensities consistent with HzRG-centred protoclusters?}

The galaxy overdensities seen here can be characterised by comparing them to predictions from cosmological simulations. \citet{saro}, for example, used hydrodynamical simulations to show that the known protocluster MRC 1138-262 is consistent with being the high--redshift progenitor of a rich galaxy cluster at $z=0$. 

\citet{chiang} follow the evolution of $\sim$ 3000 clusters in the Millennium Simulation \citep{springel,guo} from the highest redshifts to $z=0$, and can therefore match the properties of a local massive cluster to the high--redshift protocluster it grew from. A cluster, in the context of their work, is defined as a gravitationally bound dark matter halo of mass $> 10^{14}$M$_{\subsun}$h$^{-1}$; a protocluster is the high--redshift progenitor of such a $z=0$ object. 

The \citet{guo} semi-analytic galaxy model (SAM) used in the cluster simulations investigated by Chiang et al., as well as many other SAMs, have been known to under-predict the number of galaxies with extremely large star--formation rates (SFR) at high redshifts. However, the {\it Herschel} results can still be compared with the simulations assuming that the simulated galaxy population with the highest rank of SFR at a given redshift is a reasonable tracer of the population of heavily star-forming {\it Herschel}--detected galaxies at those redshifts. Because of the fixed flux density limit of our Herschel survey, the number density of {\it Herschel} SMGs that is detected is a strong function of the redshift, which translates into a limiting SFR threshold that increases with redshift.

In order to ensure that the approximate shape of this redshift-selection function is also present in the simulation, a redshift-dependent selection on SFR is applied that mimics that in the real data. This is accomplished by requiring that $SFR \geq c \times 10^{0.2z}$ M$_{\subsun}$ yr$^{-1}$. The $z$ dependence of this cut accounts for the increase in the SFR detection limit with redshift (e.g. see Yun et al. 2001). The normalisation constant, $c$, is adjusted such that the mean surface density of simulated sources matches that observed in the reference field. The analysis is carried out for protoclusters located at $z=2.5$, halfway between the redshifts of the two HzRG fields with the strongest overdensities, for which a value of $c = 17.6$ is appropriate. The redshift range searched is the same as that expected for the colour selection applied to the real data: $\Delta z =\pm 0.2(1+z) = 0.7$. It should be noted that the degree of fore- and background contamination and the size of the simulated overdensities produced using this selection do not depend strongly on the form of the SFR selection function. However, the results are sensitive to the uncertainties in the redshift range and source surface density associated with the {\it Herschel} colour selection, which are assumed here to be correct in the absence of more accurate data.

\begin{figure}
\centering
\includegraphics[scale=0.35, angle=-90]{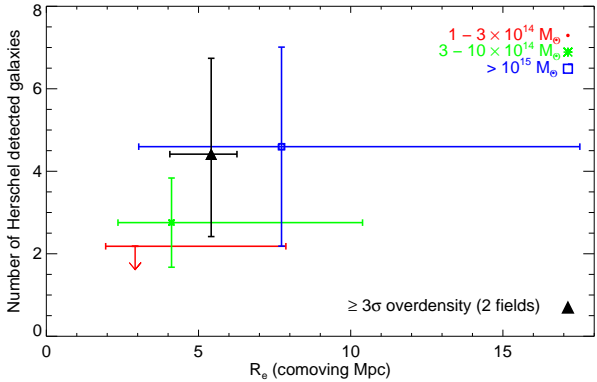}
\caption{\protect\label{size_plot} The number of galaxies contained within simulated protoclusters as a function of their radial extent, for 3 different bins in descendent (i.e. $z=0$) mass (coloured points). The radial extent here is given in terms of the effective radius, $R_{e}$, which encloses $\sim$ 60\% of the total mass of the protocluster. The black points show the size (adjusted to $R_{e}$) and number of galaxies (measured at $R_{e}$) of the average $\geq 3\sigma$ overdensities in the {\it Herschel} sample, determined from the maximum of the radial extent calculated in Section \protect\ref{rad_extent_sec}. The uncertainties on the simulated points represent the spread in the simulated distribution; the uncertainties on the real data points come from the spread in the radial extent calculation.}
\end{figure}

The most massive ($>10^{15}$M$_{\subsun}$) simulated protoclusters are also generally larger in radial extent and contain more galaxies than their low--mass counterparts. Figure \ref{size_plot} illustrates this trend for three separate bins in cluster mass defined at $z=0$. Also shown are the positions of the average {\it Herschel} $\geq 3\sigma$ galaxy overdensities, taken from the radial extent calculation in Section \ref{rad_extent_sec}. These imply an enclosed mass for these structures of $>10^{14}$ M$_{\subsun}$, which is consistent with that determined previously for HzRG--selected protoclusters \citep[e.g.][]{venemans2007,hatch2011a}

This is a convincing picture, but the large redshift range searched means that the galaxies contributing to the overdensity may lie a long way from the HzRG in redshift space. Investigating this with the simulations implies that overdensities of similar strength to those seen here (e.g. $\rho_{g} \sim 1.7$)  typically consist of between 1 and 4 protocluster structures along the line of sight, within the volume defined by the search radius and redshift uncertainty. Of these, the largest protocluster in the region contributes $\sim$3 galaxies on average to the overall source numbers. Whilst this is consistent with the $\sim$5 excess {\it Herschel} galaxies, it is clear that the redshift uncertainty must be reduced to confirm the nature of the galaxy excesses seen here.

\subsection{Comparison with previous protocluster searches}

Of the 26 HzRG fields targeted with {\it Herschel} 8 have previously been identified as containing a rich protocluster (references given in Table \ref{obs_table}). A further 3 have significant overdensities of either Spitzer--selected galaxies \citep[PKS 0529-549 \& 4C23.56, both cirrus--contaminated here;][]{galametz, mayo} or near--infrared selected galaxies \citep[MG 2308+0336;][]{hatch2011a}. For comparison, the 2 fields found here to contain $>3\sigma$ far--infrared overdensities of galaxies (MP J1758-6738 and  NVSS J111921-363139) also show overdensities of Spitzer--selected galaxies of 2.6 and 3$\sigma$ respectively \citep{wyl}.

One field, PKS 1338-1942, has previously been found to have an overdensity at 1.2 mm, through a combination of sub--millimetre, radio and optical imaging \citep{breuck}. 
This result is interesting as none of the spectroscopically confirmed Ly$\alpha$ emitters used to originally identify it as a rich protocluster \citep{venemans2007} were detected at 1.2 mm, suggesting that there is no overlap between them and the dusty galaxy population.  This is consistent with the lack of significant overdensities in the known protoclusters considered here. It should be noted that the majority of the sources contributing to the overdensity are detected in the SPIRE data, but the low resolution means that many are blended. In addition the photometric redshift constraints at 1.2 mm are weak which, combined with the uncertainty in the far--infrared colour redshift selection used here, means that no corresponding overdensity is seen at these wavelengths.

The HzRG sample used here, and that used for the linked Spitzer studies of \citeauthor{galametz} (3.6 and 4.5\mic) and \citeauthor{mayo} (24\mic) have 10 of the low--cirrus HzRG fields in common (16 in total). Six of these overlaps are known, spectroscopically confirmed, protoclusters, 4 of which do not show galaxy overdensities in either the Spitzer or the {\it Herschel} data. In total only 25\% of the overlapping fields are overdense at 3.6, 4.5 or 24\mic. 

Eight fields also overlap with the CARLA survey \citep{wyl}, which searches for galaxy overdensities in deeper Spitzer 3.6 and 4.5\mic data than that used by \citet{galametz}. Of these, only one is a known protocluster, and it is not significantly overdense in either sample. In total 50\% of the overlapping, low--cirrus, fields show  $>3\sigma$ Spitzer overdensities.

A potential issue here is the assumption that any protocluster overdensity present is centred on the HzRG position. \citet{hayashi} studied the distribution of $H\alpha$ emitters in the vicinity of a protocluster, and found they lay in three distinct clumps, one of which surrounded a HzRG. Visual inspection of the spatial distribution of colour--selected galaxies (Figure \ref{radecfig}) suggests that this may be the case for the MRC 0052-241 field, which contains a clump of sources located just outside the edge of the search radius. 

\section{Summary and Conclusions}
\label{conclude}

This paper has analysed the far--infrared environments surrounding 19 HzRGs, after excluding an additional 7 fields for post--background subtraction cirrus--contamination, which can manifest as a spurious protocluster signal. 

Simple source counting finds that, on average, at 500\mic, the fields have a higher galaxy surface density than the background within the central 6 comoving Mpc. Improved statistical constraints on this result are obtained by restricting the analysis to potential protocluster members only (selected using a far--infrared colour cut); this reveals significant overdensities in 2 fields. Comparison with simulated protoclusters drawn from the Millennium Simulation shows that these overdensities are consistent with protoclusters of mass $>10^{14}$ M$_{\subsun}$, but that it is also possible that the excess galaxies come from several low--mass protocluster structures located within the volume searched.  

The generally low observed surface density of these objects at these wavelengths, which must in part be ascribed to the large SPIRE beam, makes overdensities particularly hard to identify here. These results demonstrate that it is possible to identify potential protocluster candidates in {\it Herschel} data, but the inefficiency of the process, and the failure to detect the known rich protoclusters in the sample makes it clear that targeting HzRGs is a less successful technique here than it has proved at other wavelengths. It seems likely that SPIRE is probing different structures than those identified using the classical narrow--band or mid--infrared imaging techniques. This conclusion is reinforced by the growing number of serendipitous discoveries, in far--infrared surveys, of protoclusters with no central HzRG (e.g. Clements et al. 2013, MNRAS submitted, and the spectroscopically confirmed cluster of starburst galaxies presented in Ivison et al. \citeyear{ivison}). Further work is needed to understand the different populations this implies. 

Future work will combine these SPIRE data with forthcoming radio, optical and near--infrared data to improve the selection of protocluster member galaxies, and expand the search for associated structures across the whole area of the fields.

\section*{Acknowledgements}

EER acknowledges financial support from NWO (grant number: NWO-TOP LOFAR 614.001.006). GDL acknowledges financial support from the European Research Council under the European  Community's  Seventh Framework Programme (FP7/2007-2013)/ERC grant agreement n. 202781. NS is the recipient of an ARC Future Fellowship. Support for this work was provided by NASA through an award issued by JPL/Caltech.

SPIRE has been developed by a consortium of institutes led by Cardiff University (UK) and including Univ. Lethbridge (Canada); NAOC (China); CEA, LAM (France); IFSI, Univ. Padua (Italy); IAC (Spain); Stockholm Observatory (Sweden); Imperial College London, RAL, UCL-MSSL, UKATC, Univ. Sussex (UK); and Caltech, JPL, NHSC, Univ. Colorado (USA). This development has been supported by national funding agencies: CSA (Canada); NAOC (China); CEA, CNES, CNRS (France); ASI (Italy); MCINN (Spain); SNSB (Sweden); STFC (UK); and NASA (USA).

\appendix

\section{The spatial distribution of the galaxies}
\label{ra_app}

\begin{figure*}
\centering
\includegraphics[scale=0.5, angle=0]{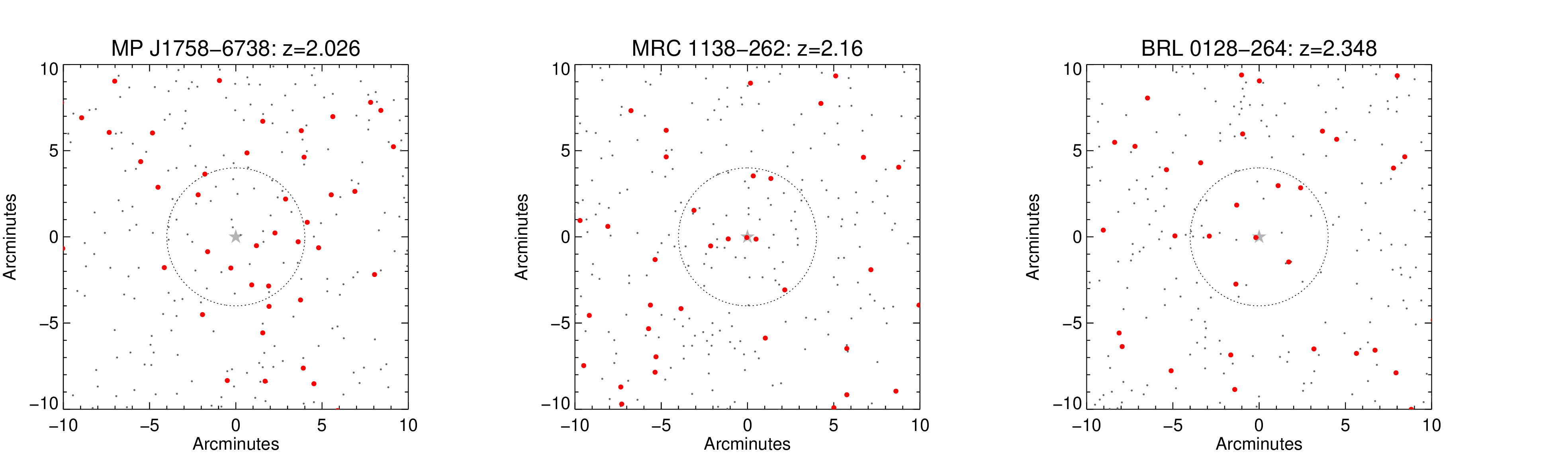}
\includegraphics[scale=0.5, angle=0]{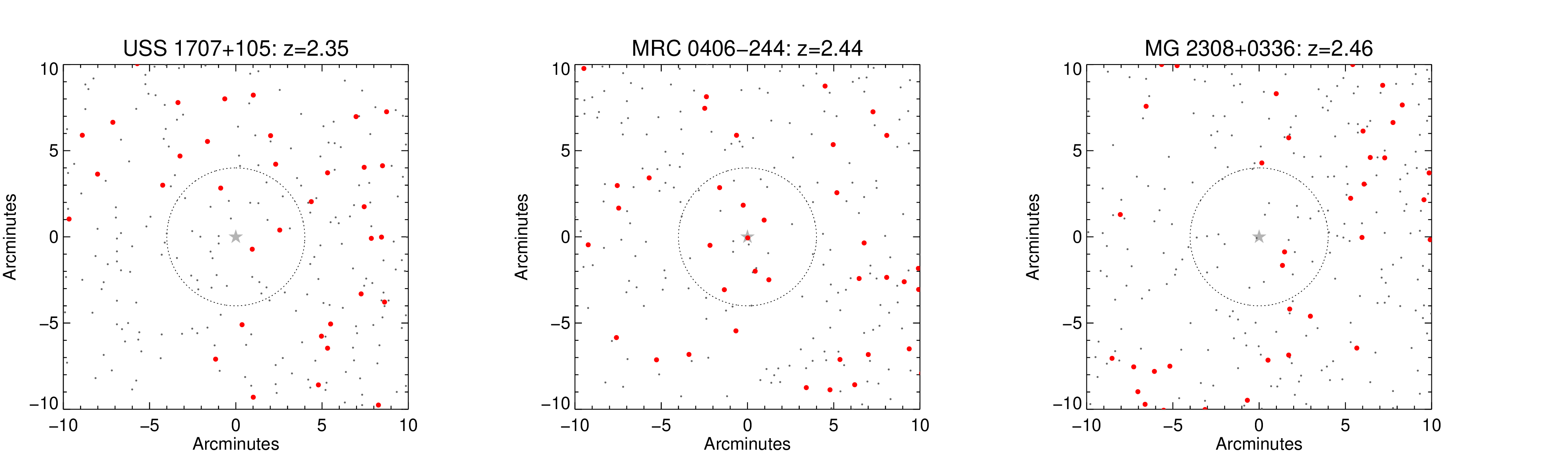}
\includegraphics[scale=0.5, angle=0]{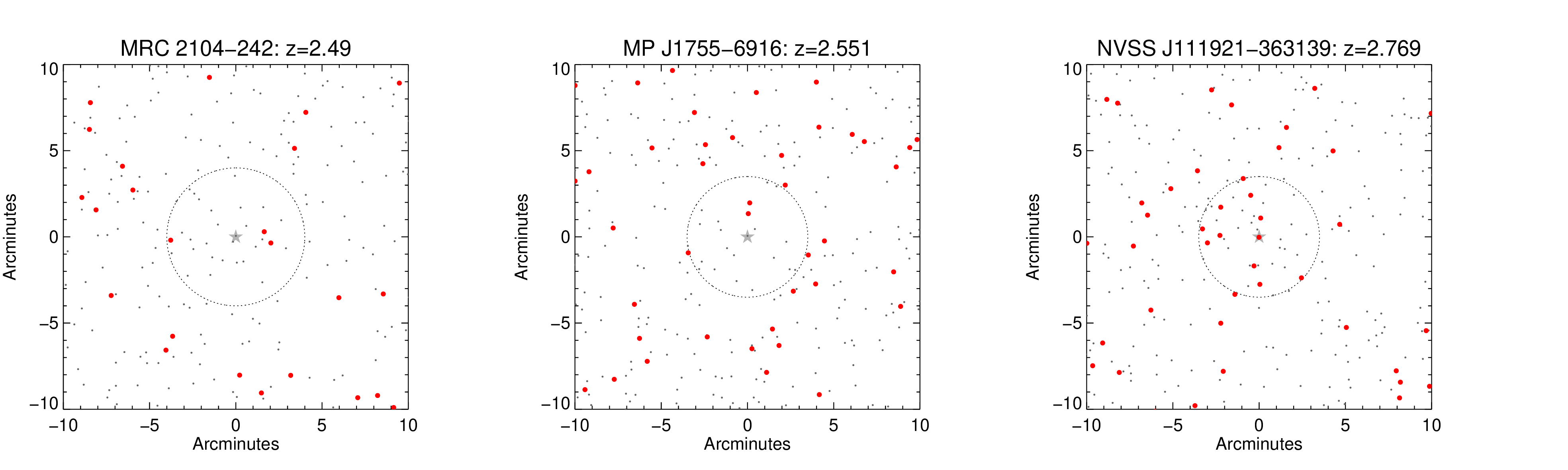}
\caption{The central spatial distribution of the galaxies in the MC500 catalogue colour--selected to lie within $\pm 0.2(1+z)$ of the HzRG (red points), as well as those $>3\sigma$ in the full 250\mic catalogue (grey points). The circle indicates the size of the 6 comoving Mpc search radius in each field.  The position of the HzRG is shown by the grey star. The 7 fields flagged as containing a high level of cirrus contamination are excluded here. \protect\label{radecfig}}
\end{figure*}

\begin{figure*}
\centering
\includegraphics[scale=0.5, angle=0]{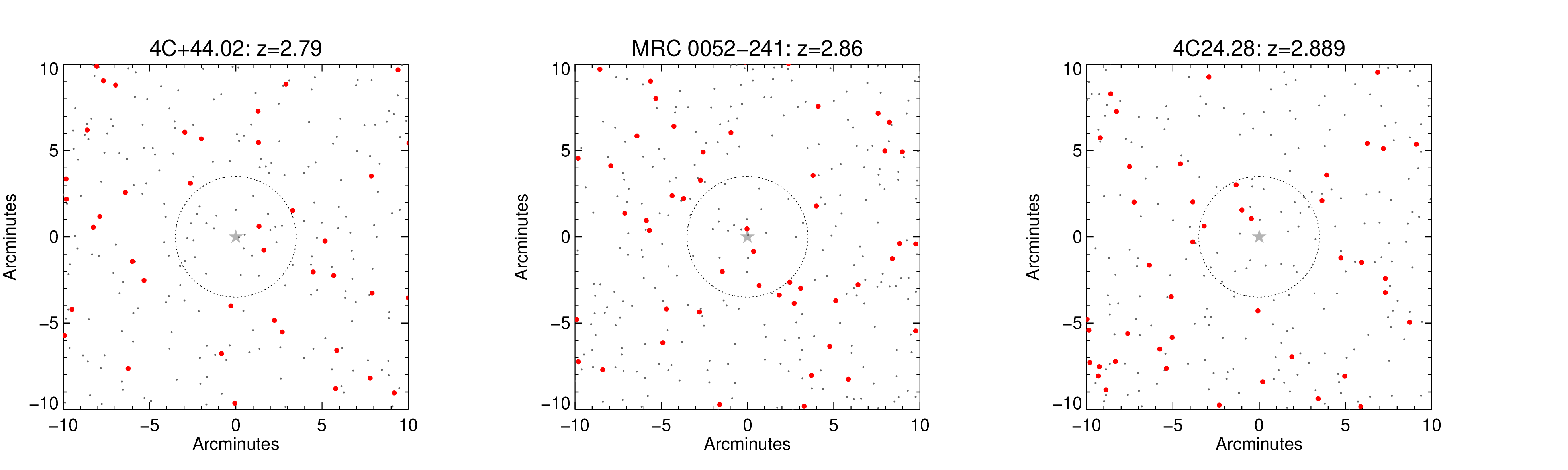}
\includegraphics[scale=0.5, angle=0]{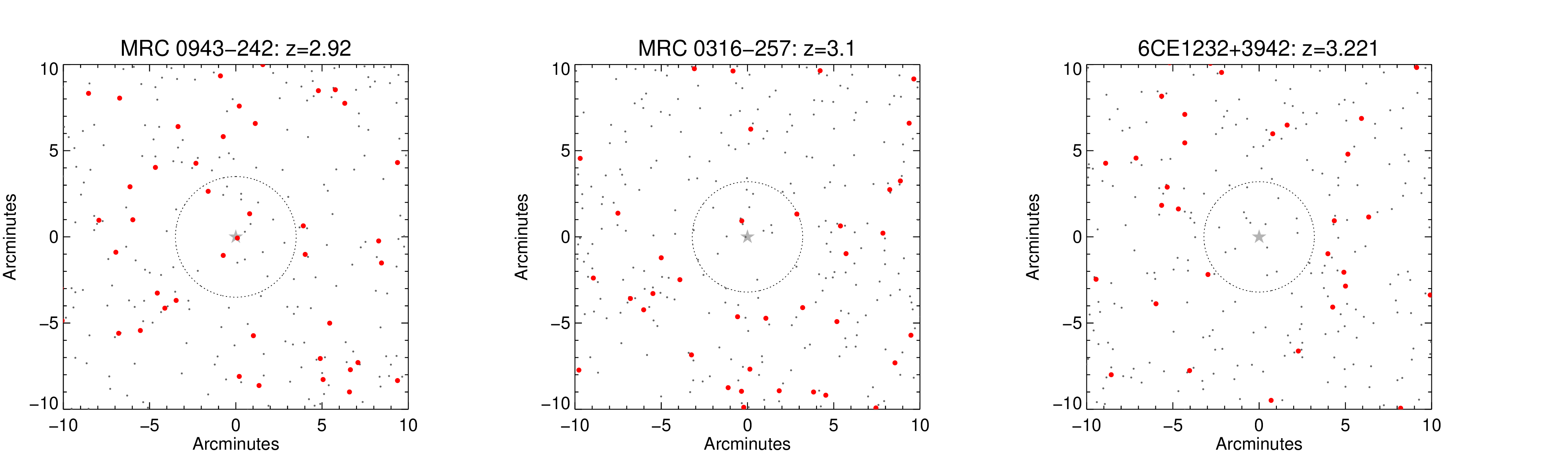}
\includegraphics[scale=0.5, angle=0]{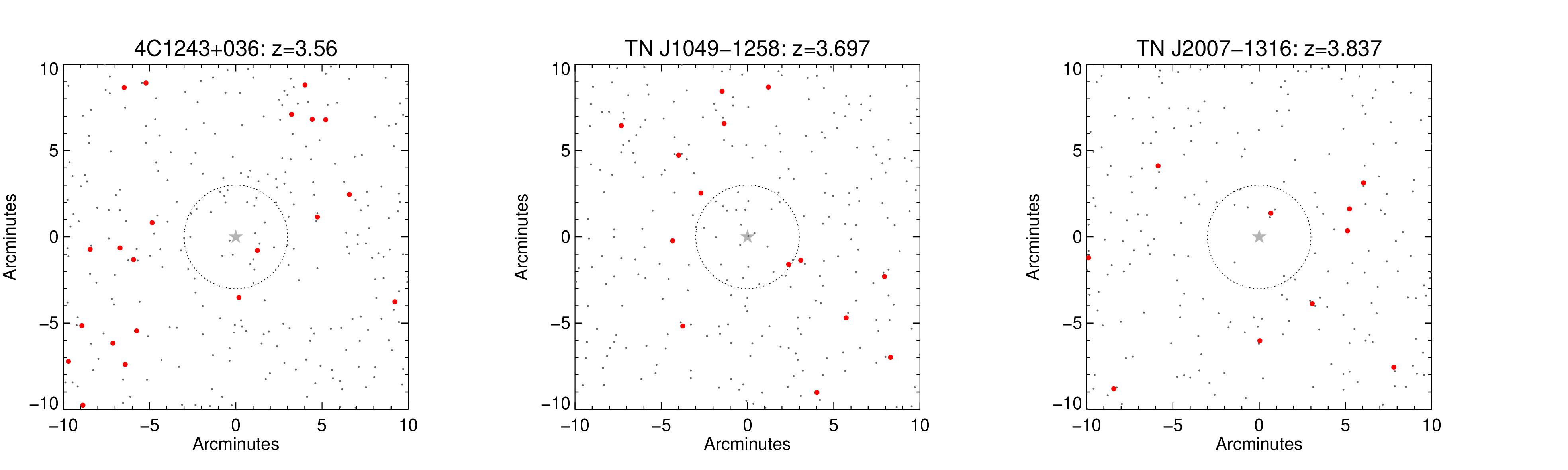}
\includegraphics[scale=0.5, angle=0]{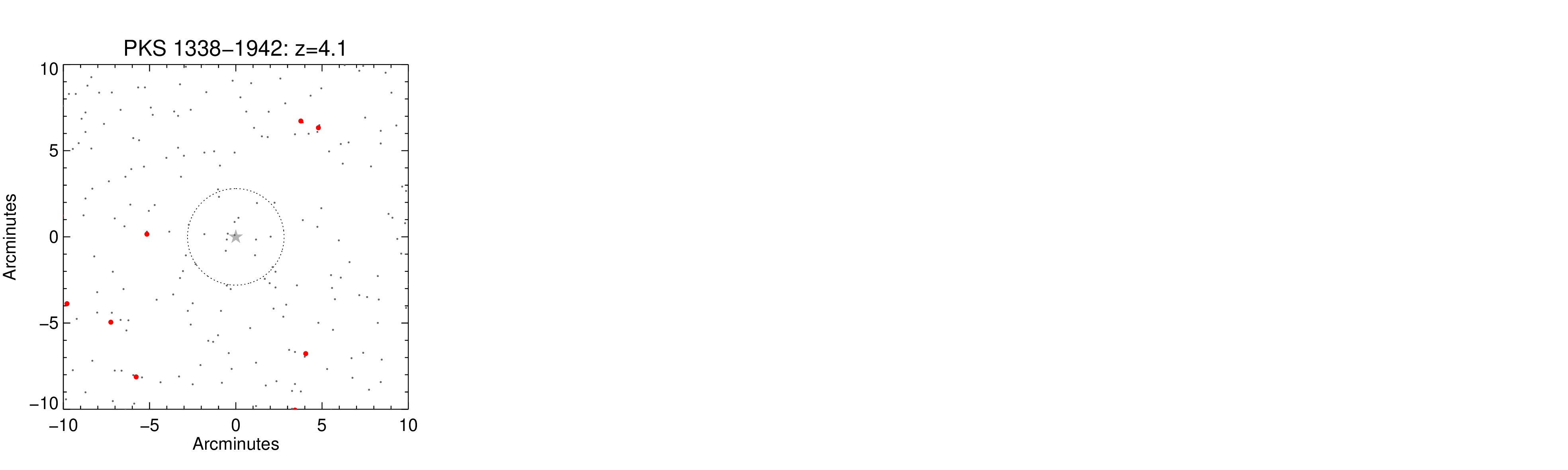}
\end{figure*}

\bsp

\label{lastpage}

\end{document}